\documentclass[acmtog, nonacm]{acmart}

\usepackage{booktabs} 
\usepackage{graphicx}
\usepackage{subcaption}
\newtheorem{claim}{Claim}

\usepackage[ruled]{algorithm2e} 

\SetAlFnt{\small}
\SetAlCapFnt{\small}
\SetAlCapNameFnt{\small}
\SetAlCapHSkip{0pt}

\copyrightyear{2025}
\acmYear{2025}

\begin{document}
\title{Flexible 3D Cage-based Deformation via Green Coordinates on Bézier Patches}

\author{Dong Xiao}
\email{xiaodong@ustc.edu.cn}
\affiliation{%
  \department{School of Mathematical Sciences}
  \institution{University of Science and Technology of China}
  \country{China}
}
\author{Renjie Chen}
\authornote{Corresponding author}
\email{renjiec@ustc.edu.cn}
\affiliation{%
  \department{School of Mathematical Sciences}
  \institution{University of Science and Technology of China}
  \country{China}
}

\begin{abstract}
Cage-based deformation is a fundamental problem in geometry processing, where a cage, a user-specified boundary of a region, is used to deform the ambient space of a given mesh. Traditional 3D cages are typically composed of triangles and quads. While quads can represent non-planar regions when their four corners are not coplanar, they form ruled surfaces with straight isoparametric curves, which limits their ability to handle curved and high-curvature deformations. In this work, we extend the cage for curved boundaries using Bézier patches, enabling flexible and high-curvature deformations with only a few control points. The higher-order structure of the Bézier patch also allows for the creation of a more compact and precise curved cage for the input model. Based on Green’s third identity, we derive the Green coordinates for the Bézier cage, achieving shape-preserving deformation with smooth surface boundaries. These coordinates are defined based on the vertex positions and normals of the Bézier control net. Given that the coordinates are approximately calculated through the Riemann summation, we propose a global projection technique to ensure that the coordinates accurately conform to the linear reproduction property. Experimental results show that our method achieves high performance in handling curved and high-curvature deformations. 
\end{abstract}

%
%


%
%


\begin{teaserfigure}
    \centering
    \includegraphics[width=\textwidth]{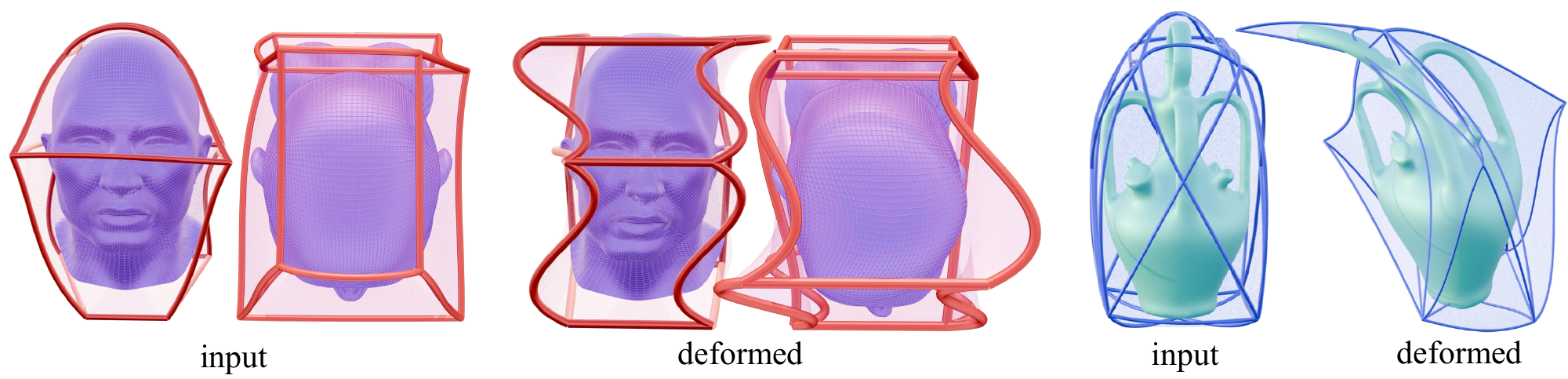}
    \caption{The high-order structure of Bézier patches (left: tensor product Bézier patches, right: Bézier triangles) allows us to design a tightly fitted cage and achieve curved deformations. }
    \label{fig:teaser}
\end{teaserfigure}

\maketitle
\section{Introduction}
Space deformation is a widely studied topic in geometry processing, focusing primarily on deforming the ambient space of an embedded shape~\citep{2009VariationalGC}. One commonly used method is cage-based deformation, where the embedded object deforms according to the changes in the cage as the user manipulates the shape from the source cage to the target cage.

Cage-based deformation mainly constructs barycentric coordinates, which form a partition of unity $\phi_{i}(\eta)$ for each point $\eta \in \mathbb{R}^3$ within the cage, satisfying $\sum_{i=1}^{N}\phi_{i}(\eta) = 1$ and $\eta=\sum_{i=1}^{N}\phi_{i}(\eta) \textbf{v}_i$, where $\textbf{v}_i$ represent the cage vertices. Various methods exist for computing these coordinates, such as Mean Value Coordinates~\citep{2005MVC, 2018MVCQuad}, Harmonic Coordinates~\citep{2007Harmonic}, and Poisson Coordinates~\citep{2013Poisson}. However, these methods do not prioritize the conformal property, which may result in shearing artifacts. To address this issue,~\citet{2008Green} introduce Green coordinates, which take into account both the cage vertices $\textbf{v}_i$ and the cage normals $\textbf{n}_j$. Green coordinates are expressed as $\eta=\sum_{i=1}^{N}\phi_{i}(\eta) \textbf{v}_i + \sum_{j=1}^{M}\psi_{j}(\eta) \textbf{n}_j$, achieving quasi-conformal mapping in 3D and exhibiting shape-preserving effects.

Besides coordinates, another key factor is the geometry of the cage structure. Earlier work always represents the cage as an oriented simplicial surface~\citep{2005MVC, 2008Green, 2009VariationalGC}, which has limited capabilities for shape control. To fill this gap,~\citet{2018MVCQuad} and~\citet{2022QuadGreen} propose mean value coordinates and Green coordinates for tri-quad cages. A quadrilateral (quad) in 3D can represent a non-planar region when its four vertices are not coplanar. However, every isoparametric curve of the quad is straight, making it challenging to control curved deformations. To adapt to high-curvature deformations, \citet{2023PolyGreen} propose polynomial cages and achieve flexible deformations for curved shapes. However, this method is only available in 2D. Moreover, S-patch \citep{2015Spatch} and $C^{0}$ GC patch \citep{2024GCpatych} can be utilized to construct barycentric coordinates for high-order structures in 3D. However, it is non-trivial for them to establish Green coordinates with shape-preserving capabilities due to the absence of cage normals.

In this research, we utilize Bézier patches to represent 3D cages and construct Green coordinates for cage-based deformation. To the best of our knowledge, this is the first work to construct coordinates with normal control for high-order cages with curved boundaries in 3D. The flexibility of our approach is demonstrated in the following two aspects: 1) The curved structure of Bézier patches enables us to design a more compact and tightly fitted cage for the input model; and 2) The smoothness of Bézier patches enables a more flexible 3D shape deformation, particularly for curved and high-curvature scenarios. 

We derive the Green coordinates for the Bézier cage based on Green's third identity, which facilitate a shape-preserving deformation. Given that establishing Green coordinates requires cage normals, our method constructs surface normals based on the Bézier control net and subsequently expresses the coordinates in terms of both the positions and normals of the control vertices. Due to the absence of a closed-form solution for the integration, we utilize Riemann summation to approximately calculate the coordinates. We then introduce a global projection method to map these approximate results onto the solution space that satisfies precise linear reproduction, ensuring the theoretical validity of the results.

Fig.~\ref{fig:teaser} shows typical examples of our approach, in which we design tightly fitted Bézier cages (left: tensor product Bézier patches, right: Bézier triangles) to achieve curved deformation. In the experimental section, we compare our method with other coordinate systems and cage representations. The results demonstrate that our approach yields smooth surface boundaries by leveraging higher-order structures and achieves superior performance in handling curved and high-curvature deformations. The source code is available at~\url{https://github.com/Submanifold/BezierGreen}.

\section{Related work}
Cage-based deformation has received considerable attention due to its broad range of applications. Given a cage, various methods can be employed to establish the coordinates, depending on the type of cage and the specific geometric requirements. In this section, we will introduce the most relevant work to our study. For a comprehensive review of different coordinate systems, one can refer to \citep{2015ReviewBarycentric} and \citep{2017HormannBarycentric}. Additionally, \citet{2024Survey} provide a comprehensive survey for cage-based deformation in 3D.

\subsection{Interpolatory coordinates}
Interpolatory coordinates mainly represent the position within a geometric shape by a set of weights that form a partition of unity. Most of these coordinates are known as Generalized Barycentric Coordinates (GBC)~\citep{2017HormannBarycentric}. They are widely utilized in cage-based deformation. Mean Value Coordinates (MVC), initially proposed for 2D and 3D polygons~\citep{2003MVCPoly, 2005MVCPoly}, have inspired numerous subsequent studies in this field. \citet{2005MVC} define MVC for 3D triangle meshes and apply this technique for cage-based deformation by projecting the cage onto a unit sphere. Building on a similar concept, \citet{2006Spherical} propose spherical barycentric coordinates for planar n-gons; \citet{2007PMVC} introduce Positive Mean Value Coordinates (PMVC), which eliminate artifacts for non-convex cages by avoiding negative weights in the original MVC. \citet{2013Poisson} introduce Poisson coordinates based on the Poisson integral formula, which can be considered an extension of the MVC framework. Additionally, several studies focus on extending MVC to nonlinear cage structures. Cubic MVC~\citep{2013CubicMVC} enables mean-value interpolation for 2D or 3D curves. QMVC~\citep{2018MVCQuad} extends MVC to non-planar 3D quads.~\citet{2024Stochastic} establish barycentric coordinates through stochastic sampling and Monte Carlo integration, which supports broad cage types.

Besides the MVC framework, other techniques have also been widely explored. \citet{2007Harmonic} construct harmonic coordinates by numerically solving the Laplace equation. Maximum Entropy Coordinates~\citep{2008MaxEntropy} and Maximum Likelihood Coordinates~\citep{2023MaximumLikelihood} conceptualize coordinates as probability distributions and construct the coordinate system using probability theory. \citet{2019GeneralBarycentric} and \citet{2023VBC} accumulate the contributions of vertex triangles, and they are able to produce positive weights for non-convex polygons.

\subsection{Coordinates with normal control}
In addition to barycentric coordinates, another important class is known as Green coordinates~\citep{2008Green}, which consider not only the vertices but also the normals. Green coordinates have a closed-form solution for triangle meshes, which is thoroughly derived in \citet{2009DriveGreen}. Many works establish Green coordinates on different cage structures. \citet{2023PolyGreen} and \citet{2024PolyHighOrder} propose polynomial 2D Green coordinates and derive the closed-form expression. \citet{2022QuadGreen} extend Green coordinates to tri-quad cages in 3D and employ robust Riemann summation for coordinate computation.

There are also other normal-controlled coordinates, such as biharmonic coordinates~\citep{2012Biharmonic, 2024Biharmonic3D} and Somigliana coordinates~\citep{2023Somigliana}. Furthermore, Cauchy coordinates~\citep{2009CBC} and polynomial Cauchy coordinates~\citep{2024PolyCauchy} constructed in the complex plane can be regarded as equivalent to Green coordinates after specific derivations. To the best of our knowledge, there are currently no Green coordinates defined on high-order cages with curved boundaries in the 3D space.

\subsection{High order structures}
Previous works on geometric modeling typically utilized piecewise linear data to represent various shapes. However, in recent years, there has been a growing emphasis on high-order structures, including Bézier triangles, Bézier tetrahedrals, and tensor product Bézier surfaces~\citep{1993CAGDBook}. Researchers such as \citet{2019TriWild}, \citet{2020BezierGuarding}, and \citet{2022Precise2DMeshing} employ diverse strategies to construct high-order meshes in the 2D domain. \citet{2021HighOrderTetrahedral} proposes an automatic algorithm to convert triangle meshes to high-order meshes in 3D. \citet{2024HighOrderShell} propose to construct high-order shells that facilitate attribute transfer.

In addition, several studies explore the application of high-order concepts to shape deformation. For instance, \citet{2015Spatch} and \citet{2024GCpatych} propose generalized barycentric coordinates using transfinite interpolation techniques, which can be extended to cage-based deformation. However, these methods usually lack normal controls, which poses challenges in achieving shape-preserving deformations.
\section{Preliminaries}
\subsection{Green coordinates for cage-based deformation}
Cage-based deformation with normal control can produce more realistic and shape-preserving results. We will first recall the Green coordinates for oriented simplicial surfaces~\citep{2008Green}. The source cage is represented as a triangle mesh with vertices $\{\textbf{v}_i\}{_{i=1}^{N}}$, faces $\{f_j\}_{j=1}^{M}$, and face normals $\{\textbf{n}_j\}{_{j=1}^{M}}$, where $N$ and $M$ represent the number of vertices and faces. Green coordinates mainly construct a series of functions $\{\phi_{i}(\eta)\}{_{i=1}^{N}}$ and $\{\psi_{j}(\eta)\} {_{j=1}^{M}}$ regarding the cage vertices and normals to describe an arbitrary position $\eta$ within the cage, which can be expressed as:
\begin{equation}
\label{eq:phi_psi}
\eta = \sum_{i=1}^{N} \phi_{i}(\eta) \textbf{v}_{i} + \sum_{j=1}^{M} \psi_{j}(\eta) \textbf{n}_{j}. 
\end{equation}
Then, we can modify the source cage according to our requirements. If the vertices and normals of the target cage are $\{\tilde{\textbf{v}}_{i}\}{_{i=1}^{N}}$ and $\{\tilde{\textbf{n}}_{i}\}{_{i=1}^{M}}$, respectively, the deformed location $\tilde{\eta}$ is calculated as:
\begin{equation}
\label{eq:new_location}
\tilde{\eta} = \sum_{i=1}^{N} \phi_{i}(\eta) \tilde{\textbf{v}}_{i} + \sum_{j=1}^{M} \psi_{j}(\eta) s_j \tilde{\textbf{n}}_{j}, 
\end{equation}
where $s_j$ is a scalar factor to maintain quasi-conformality, whose computation method is detailed in \citet{2008Green}.

The construction of Green coordinates ($i.e.$, the calculation of $\phi_{i}(\eta)$ and $\psi_{j}(\eta)$ in Eq.~\eqref{eq:phi_psi}) is mainly based on Green's third identity. If $\Omega$ is a bounded region in $\mathbb{R}^{3}$ and function $f$ is harmonic, for an arbitrary $\eta \in \Omega$, $f(\eta)$ can be expressed as a boundary integral:
\begin{equation}
\label{eq:1}
f(\eta) = \underbrace{\int_{\xi \in \partial \Omega}{f(\xi)\frac{\partial G}{\partial n}(\xi, \eta) \ \mathrm{d}\xi}}_{f_{D}(\eta)}  \underbrace{-\int_{\xi \in \partial \Omega}{G(\xi, \eta) \frac{\partial f}{\partial n}(\xi) \ \mathrm{d}\xi}}_{f_{N}(\eta)},
\end{equation}
where $G$ is the fundamental solutions of the Laplace equation $\Delta G_{\xi}(\xi, \eta) = \delta(||\xi - \eta||)$. In 3D, $G$ has the following expression: 
\begin{equation}
\label{eq:2}
G(\xi, \eta)=-\frac{1}{4\pi||\xi - \eta||}. 
\end{equation}
The directional derivative of $G$ with respect to $\xi$ can be expressed as:
\begin{equation}
\label{eq:pgpn}
\frac{\partial G}{\partial n}(\xi, \eta) = \frac{(\xi - \eta) \cdot \textbf{n}_{\xi}}{4\pi ||\xi - \eta||^{3}}, 
\end{equation}
where $\textbf{n}_{\xi}$ is the normal at $\xi$.

Eq.~\eqref{eq:1} gives a continuous representation of $f$ as a boundary integral. When the cage is a triangle mesh, we want $f$ to be piecewise linear on it. \citet{2008Green} propose the following expression for $f$:
\begin{equation}
\label{eq:2_2}
f(\xi) = \sum_{i \in \mathrm{Ngbr}(\xi)}\Gamma^{i}(\xi)\textbf{v}_i, 
\end{equation}
where $\Gamma^{i}(\xi)$ is a hat function that takes the value $1$ at $\textbf{v}_i$, $0$ at all other vertices, and varies linearly across each face of its 1-ring neighbour. Then, the coordinates $\phi_{i}(\eta)$ and $\psi_{j}(\eta)$ can be calculated as:
\begin{equation}
\begin{alignedat}{2}
\phi_{i}(\eta) &= \int_{\xi \in F_1(i)}{\Gamma^{i}(\xi)\frac{\partial G}{\partial n}(\xi, \eta) \ \mathrm{d}\xi}, \quad &&
 \psi_{j}(\eta) &= \int_{\xi \in t_j}{-G(\xi, \eta) \ \mathrm{d}\xi},
\end{alignedat}
\label{eq:phi_psi_Lipman}
\end{equation}
where $F_{1}(i)$ denotes the 1-ring neighbour of $\textbf{v}_i$.

While \citet{2008Green} originally define Green coordinates on triangle meshes, \citet{2022QuadGreen} extend this technique to quad meshes. A quad $\textbf{q}(u, v)$ is represented as a bilinear interpolation surface of four vertices $\{\textbf{q}_i\}_{i=0}^{3}$, governed by the following parametric equation:
\begin{equation}
\label{eq:q_uv}
\textbf{q}(u, v) = (1-u)(1-v)\textbf{q}_0 + u(1-v) \textbf{q}_1 + uv \textbf{q}_2 + v(1-u) \textbf{q}_3.
\end{equation}
A quad can define a non-planar ruled surface, where the isoparametric curves $\textbf{q}(u_0, v)$ and $\textbf{q}(u, v_0)$ are straight. Inspired by the polynomial Green coordinates proposed by \citet{2023PolyGreen} in 2D, which construct the cage with polynomial curves as $f(t) = \sum_{i} c_i t^{i}$, we propose to extend the 3D cage with higher-order structures.

\subsection{Tensor product Bézier surfaces}
Triangle meshes are limited to first-order approximations due to their piecewise linearity, which restricts their ability to accurately represent continuous surfaces. To address this limitation, previous studies have explored the use of high-order geometric representations to enhance surface expression. A Bézier patch is typically constructed as a high-order parametric surface from a series of control points. Our method is capable of handling various cage types including both tensor product Bézier surfaces and Bézier triangles. In the main text, we mainly focus on tensor product Bézier surfaces, while detailed discussions on Bézier triangles are presented in Appendix A of the supplementary material.

A tensor product Bézier patch is governed by two parameters $u$ and $v$, constrained by $0 \le u \le 1, 0 \le v \le 1$. which can be understood as moving a B\'{e}zier curve alone another Bézier curve. A degree-$(m, n)$ Bézier patch $\textbf{b}(u, v)$ has $(m+1)\times(n+1)$ control points $\textbf{b}_{ij}, i \in \{0,1,2,...,m\}, j \in \{0,1,2,...,n\}$. The expression for the Bézier patch is given by:
\begin{equation}
\begin{aligned}
\label{eq:s_uv}
\textbf{b}(u,v)=\sum_{i=0}^{m}{\sum_{j=0}^{n}{\frac{m!}{i!(m-i)!}\frac{n!}{j!(n-j)!} u^{i}(1-u)^{m-i}v^{j}(1-v)^{n-j}}\textbf{b}_{ij}}.
\end{aligned}
\end{equation}
We can denote $\lambda^{ij}(u,v)=B^{m}_{i}(u)B^{n}_{j}(v)=\frac{m!}{i!(m-i)!}\frac{n!}{j!(n-j)!} u^{i}(1-u)^{m-i}v^{j}(1-v)^{n-j}$, 
which is the product of two Bernstein polynomials. Consequently,  the following expression holds:
\begin{equation}
\label{eq:abb_buvsuv}
\textbf{b}(u, v) = \sum_{i=0}^{m}{\sum_{j=0}^{n}{\lambda^{ij}(u,v) \textbf{b}_{ij}}}.
\end{equation}

\section{Method}

\subsection{Green coordinates for Bézier cage}
In this section, we derive the Green coordinates for the Bézier cage composed of tensor product Bézier patches. The coordinates are constructed based on Green's third identity (Eq.~\eqref{eq:1}). We represent the Dirichlet term as $f_D(\eta)$ and the Neumann term as $f_N(\eta)$, where $f(\eta) = f_D(\eta) + f_N(\eta)$. Let $f(\eta) = \eta$, then $f$ is harmonic. We can obtain a continuous representation of $\eta$ as a boundary integral. We directly derive the deformed position $\tilde{\eta}$ within a continuous formulation. In the following, we will denote the values after deformation with tilde lines $(e.g., \tilde{\textbf{b}}(u, v), \tilde{f}_{D}^{Q}(\eta), \tilde{\eta})$, while those before deformation remain without tilde lines $(e.g., \textbf{b}(u, v), \textbf{b}_{u}(u, v), \eta$). The Dirichlet term $\tilde{f}_{D}^{Q}(\eta)$ and Neumann term $\tilde{f}_{N}^{Q}(\eta)$ of a single Bézier patch $Q$ can be expresses as follows:
\begin{equation}
\label{eq:f_D}
\tilde{f}_{D}^{Q}(\eta) = \int_{\xi \in Q}{\tilde{f}(\xi)} \frac{\partial G}{\partial n}(\xi, \eta) \ \mathrm{d}\xi,
\end{equation}
where
\begin{gather}
\label{eq:f_D_parts}
\tilde{f}(\xi) = \tilde{\textbf{b}}(u, v), \\
\frac{\partial G}{\partial n}(\xi, \eta) = \frac{\textbf{b}(u, v) - \eta}{4\pi ||\textbf{b}(u ,v) - \eta||^{3}} \cdot \frac{\textbf{b}_u (u, v) \times \textbf{b}_v(u ,v)}{||\textbf{b}_u (u, v) \times \textbf{b}_v(u ,v)||}, \\
\mathrm{d}\xi = ||\textbf{b}_u (u, v) \times \textbf{b}_v(u ,v)|| \ \mathrm{d}u \mathrm{d}v.
\end{gather}
Additionally,
\begin{equation}
\label{eq:f_N} 
\tilde{f}_{N}^{Q}(\eta)= - \int_{\xi \in Q}{G(\xi, \eta)\frac{\partial \tilde{f}}{\partial n}(\xi) s(\xi) \ \mathrm{d}\xi},
\end{equation}
where
\begin{gather}
G(\xi, \eta) = -\frac{1}{4\pi||\textbf{b}(u, v) - \eta||}, \\
\frac{\partial \tilde{f}}{\partial n}(\xi) = \frac{\tilde{\textbf{b}}_u(u, v) \times \tilde{\textbf{b}}_v(u, v)}{||\tilde{\textbf{b}}_u(u, v) \times \tilde{\textbf{b}}_v(u, v)||}, \\
s(\xi)=\frac{||\tilde{\textbf{b}}_u(u, v) \times \tilde{\textbf{b}}_v(u, v)||}{||\textbf{b}_u(u, v) \times \textbf{b}_v(u, v)||}\label{eq:s_xi_eta}, \\
\mathrm{d}\xi = ||\textbf{b}_u (u, v) \times \textbf{b}_v(u ,v)|| \ \mathrm{d}u \mathrm{d}v.
\end{gather}
The term $s(\xi)$ (Eq.~\eqref{eq:s_xi_eta}) serves as the area-based scale factor in 3D~\citep{2022QuadGreen}.
Then, we can obtain the expressions for $\tilde{f}_{D}^{Q}(\eta)$ and $\tilde{f}_{N}^{Q}(\eta)$ as follows:
\begin{gather}
\tilde{f}_{D}^{Q}(\eta) = \iint_{u,v=0}^{1}{\frac{(\textbf{b}(u,v)-\eta) \cdot \textbf{N}(u, v)}{4\pi||\textbf{b}(u,v)-\eta||^{3}}} \tilde{\textbf{b}}(u, v)\ \mathrm{d}u \mathrm{d}v,\label{eq:f_D_concrete}
\\
\tilde{f}_{N}^{Q}(\eta) = \iint_{u,v=0}^{1}{\frac{1}{4\pi||\textbf{b}(u,v)-\eta||}} (\tilde{\textbf{b}}_{u}(u, v) \times \tilde{\textbf{b}}_{v}(u, v))\ \mathrm{d}u \mathrm{d}v.\label{eq:f_N_concrete}
\end{gather}

Although the parameters $u$ and $v$ are of high degree in $\tilde{\textbf{b}}(u, v)$, $\tilde{\textbf{b}}(u, v)$ is linear with respect to the positions of control points $\tilde{\textbf{b}}_{ij}$. Consequently, we can establish the Dirichlet term in relation to the Bézier control net. However, normals are also necessary for constructing the Neumann term. For triangle meshes, planar face normals can be calculated and applied directly. However, for high-order Bézier patches, it is non-trivial to directly represent the (unnormalized) normal $(\tilde{\textbf{b}}_u \times \tilde{\textbf{b}}_v)(u, v)$ using the normals of the control net. We can choose to construct the Neumann term based on the cross-product of any two control points. However, this will increase the number of Neumann terms from $(m+1)(n+1)$ to $[(m+1)^{2}(n+1)^{2}-(m+1)(n+1)]/2$. This increases the time required for the deformation but generates results very similar to those of the method introduced below. We will elaborate on the cross-product Neumann term in Appendix D. We notice that some studies propose approximating the normal of a Bézier patch by interpolating the normals of the control vertices and substituting these normals into the parametric formulation~\citep{2001PN, 2002FacetBasedSF}. This inspires us to apply a similar approach to approximate $(\tilde{\textbf{b}}_u \times \tilde{\textbf{b}}_v)(u, v)$ for the tensor product Bézier patch. In the following text, we will frequently need unnormalized normals, and we will no longer explicitly mention the term ``unnormalized''.

\begin{figure}[htb]
  \centering
  \includegraphics[width=0.9\linewidth]{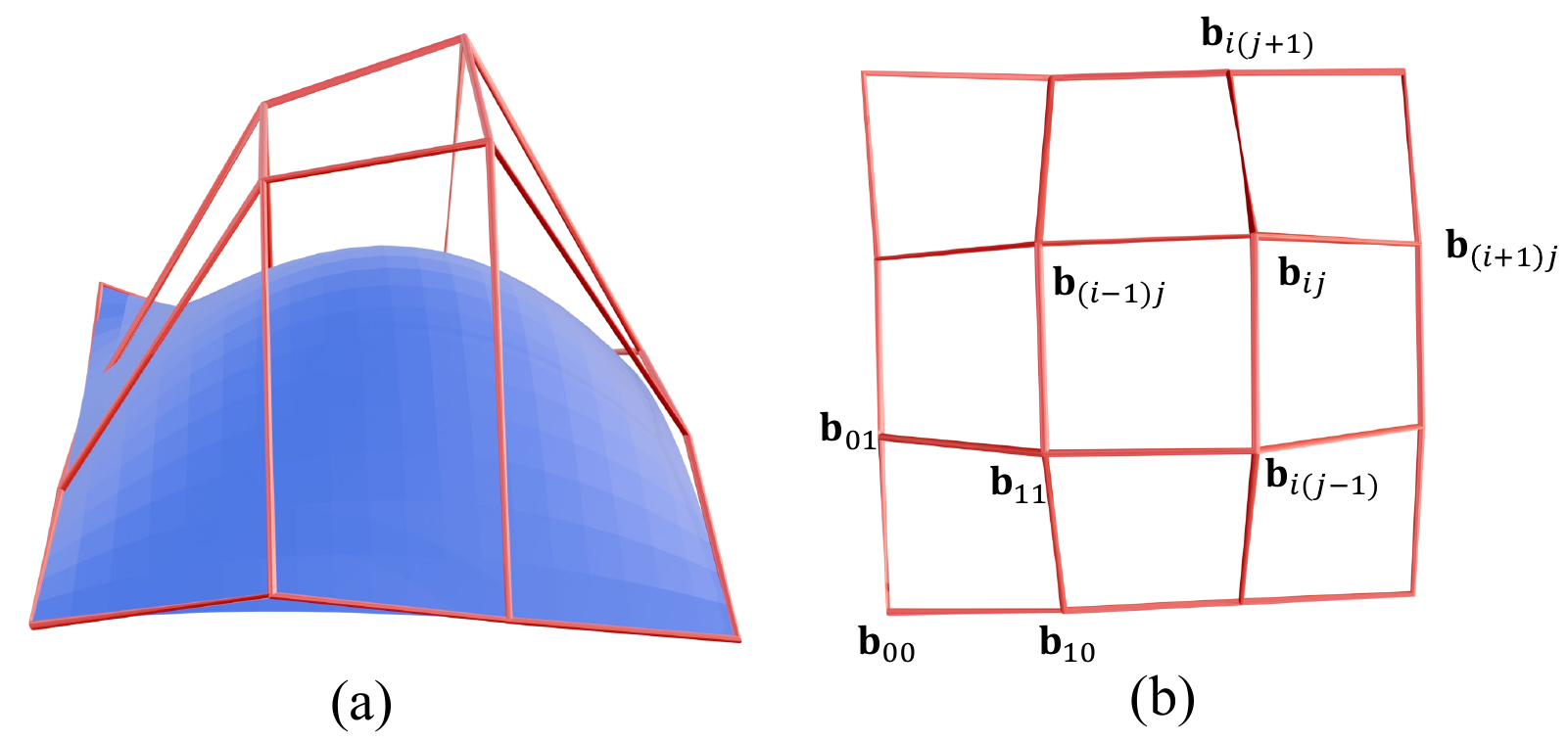}
  \caption{\label{fig:Figure7}
          Illustration of a tensor product Bézier patch (a) and its control net (b).}
\end{figure}

Given a degree-$(m, n)$ tensor product Bézier patch with control net $\textbf{b}_{ij}$, where $i \in \{0,1,2,...,m\}, j \in \{0,1,2,...,n\}$, we first compute the vertex normal $\textbf{N}_{ij}$ for the point $\textbf{b}_{ij}$. We take the normal $\textbf{N}_{00}$ at a corner in Fig.~\ref{fig:Figure7} (b) as an example. The control net is piecewise bilinear at the quad $(\textbf{b}_{00},\textbf{b}_{10},\textbf{b}_{11},\textbf{b}_{01})$, which has following parametric equation: 
\begin{equation}
\label{eq:q_{00}}
\begin{aligned}
\textbf{q}^{00}(u, v) = &(1-mu)(1-nv)\textbf{b}_{00}+mu(1-nv)\textbf{b}_{10} \\
& +mn\cdot uv\textbf{b}_{11}+nv(1-mu)\textbf{b}_{01}, u\in[0, \frac{1}{m}], v\in[0, \frac{1}{n}].
\end{aligned}
\end{equation}
Therefore, $\textbf{N}_{00}$ can be calculated as:
\begin{equation}
\label{eq:n_00}
\textbf{N}_{00}=(\textbf{q}_{u}^{00} \times \textbf{q}_{v}^{00})|_{\{u=v=0\}}=mn[(\textbf{b}_{10} - \textbf{b}_{00}) \times (\textbf{b}_{01} - \textbf{b}_{00})].
\end{equation}
For the interior control point $\textbf{b}_{ij}$, its normal $\textbf{N}_{ij}$ can be calculated by the average of its 1-ring neighbour at the control net using the following formulation:
\begin{equation}
\label{eq:n_ij}
\begin{aligned}
\textbf{N}_{ij}=\frac{mn}{4} &[(\textbf{b}_{i(j+1)} - \textbf{b}_{ij}) \times (\textbf{b}_{(i-1)j} - \textbf{b}_{ij}) \\
&+ (\textbf{b}_{(i-1)j} - \textbf{b}_{ij}) \times (\textbf{b}_{i(j-1)} - \textbf{b}_{ij}) \\
&+ (\textbf{b}_{i(j-1)} - \textbf{b}_{ij}) \times (\textbf{b}_{(i+1)j} - \textbf{b}_{ij}) \\
&+ (\textbf{b}_{(i+1)j} - \textbf{b}_{ij}) \times (\textbf{b}_{i(j+1)} - \textbf{b}_{ij})].
\end{aligned}
\end{equation}
After determining the normals of the control vertices, we can approximate the surface normal $\textbf{N}(u, v)$ for arbitrary $(u, v)$ of the Bézier patch using the following expression:
\begin{equation}
\label{eq:n_bezier_point}
\textbf{N}(u, v) = \sum_{i=0}^{m}{\sum_{j=0}^{n}{\lambda^{ij}(u,v) \textbf{N}_{ij}}},
\end{equation}
$\lambda^{ij}(u, v)$ is the Bernstein basis function of the Bézier patch, which is identical to Eq.~\eqref{eq:abb_buvsuv}. $\textbf{N}(u, v)$ approximates $(\textbf{b}_u \times \textbf{b}_v)(u, v)$ and is linearly related to $\textbf{N}_{ij}$. This linear relationship facilitates the establishment of the Neumann term based on the normals of the control vertices.

For the Dirichlet term $\tilde{f}_{D}^{Q}(\eta)$ in Eq.~\eqref{eq:f_D_concrete}, we replace $\tilde{\textbf{b}}(u, v)$ with the parametric equation~\eqref{eq:abb_buvsuv}. For the Neumann term $\tilde{f}_{N}^{Q}(\eta)$ in Eq.~\eqref{eq:f_N_concrete}, we approximate the cross product $(\tilde{\textbf{b}}_u (u, v) \times \tilde{\textbf{b}}_v(u, v))$ as $\tilde{\textbf{N}}(u, v)$ from Eq.~\eqref{eq:n_bezier_point}. After rearranging the formulas and considering all patches of the Bézier cage, the deformed position $\tilde{\eta}$ can be computed as follows:
\begin{equation}
\label{eq:phi_psi_bezier}
\tilde{\eta} = \sum_{Q}{(\tilde{f}_{D}^{Q}(\eta) + \tilde{f}_{N}^{Q}(\eta))} = \sum_{Q}{\sum_{i=0}^{m}{\sum_{j=0}^{n}{{(\phi_{Q}^{ij}(\eta) \tilde{\textbf{b}}_{ij}^{Q} + \psi_{Q}^{ij}(\eta) \tilde{\textbf{N}}_{ij}^{Q})}}}},
\end{equation}
where
\begin{equation}
\label{eq:phi_bezier}
\phi_{Q}^{ij}(\eta) = \iint_{u,v=0}^{1}{\frac{\lambda^{ij}(u ,v)(\textbf{b}_{Q}(u,v)-\eta) \cdot \textbf{N}_{Q}(u, v)}{4\pi||\textbf{b}_{Q}(u,v)-\eta||^{3}}} \ \mathrm{d}u \mathrm{d}v,
\end{equation}
\begin{equation}
\label{eq:psi_bezier}
\psi_{Q}^{ij}(\eta) = \iint_{u,v=0}^{1}{\frac{\lambda^{ij}(u, v)}{4\pi||\textbf{b}_{Q}(u,v)-\eta||}} \ \mathrm{d}u \mathrm{d}v.
\end{equation}
In above equations, $Q$ represents a single patch of the Bézier cage, $\tilde{\textbf{b}}_{ij}^{Q}$ and $\tilde{\textbf{N}}_{ij}^{Q}$ denote the positions and normals of the control net of $Q$ in the deformed cage. $\textbf{b}_{Q}(u, v)$ is the parametric equation of the source cage, $\lambda^{ij}(u, v)$ is the Bernstein coefficient in Eq.~\eqref{eq:abb_buvsuv}, and $\textbf{N}_{Q}(u, v)$ denotes the normal of the source cage. This normal will be incorporated into the solid angle calculation in the subsequent step.

\subsection{Approximated calculations for $\phi$ and $\psi$}\label{sec:4_2}
In polynomial 2D Green coordinates~\citep{2023PolyGreen, 2024PolyHighOrder}, a closed-form expression can be derived because the integration involves a rational function of a single variable. However, in the 3D case, the denominators of $\phi$ (Eq.~\eqref{eq:phi_bezier}) and $\psi$ (Eq.~\eqref{eq:psi_bezier}) are the $3/2$-th and $1/2$-th powers of a polynomial in two variables. Consequently, to the best of our knowledge, obtaining a closed-form solution is challenging. Inspired by the Riemann summation approach proposed by \citet{2022QuadGreen}, we discretize the integral region $u \in [0, 1], v \in [0, 1]$ of the Bézier patch $Q$ into smaller triangle elements $t$ and approximate the coordinates using Riemann summation. Although this approximation introduces some error (typically, the error remains minor), we develop a global projection technique to achieve precise linear reproduction, which will be elaborated upon in Section~\ref{sec:4_3}.

The denominator of $\phi$ is a cubic function of $(\textbf{b}_{Q}(u,v)-\eta)$, which suggests that we need to make finer divisions in areas closer to $\eta$ to achieve more accurate calculations. Therefore, we first project $\eta$ onto the Bézier patch $Q$ and obtain the parameter of the closest point $(u_{Q}, v_{Q})(\eta)$. This constitutes a point inversion problem~\citep{2005Projection}, and we employ gradient descent method to optimize the point-to-surface distance. Subsequently, we employ a UV pattern proposed by~\citet{2022QuadGreen} to tessellate the parameter space $[0, 1] \times [0, 1]$, ensuring dense subdivision in regions closer to $\eta$. After decomposing the integration domain into small triangles (denotes as $t$), the coordinates of the Dirichlet term $\phi_{Q}^{ij}(\eta)$ of Eq.~\eqref{eq:phi_bezier} can be approximately calculated as:
\begin{equation}
\begin{aligned}
\label{eq:phi_bezier_tessellation}
\phi_{Q}^{ij}(\eta) = &\sum_{t}{\iint_{(u, v) \in t}{\frac{\lambda^{ij}(u, v)}{4\pi} \overbrace{\frac{(\textbf{b}_{t}(u,v)-\eta) \cdot \textbf{N}_{t}(u, v)}{||\textbf{b}_{t}(u,v)-\eta||^{3}}}^{\omega_{t}(\eta)}}}  \ \mathrm{d}u \mathrm{d}v \\
\approx &\sum_{t}{\frac{\lambda^{ij}(u_t ,v_t) }{4\pi} \omega_{t}(\eta)},
\end{aligned}
\end{equation}
where $\lambda^{ij}(u_t ,v_t)$ represent the function value of $\lambda^{ij}$ at $(u_t ,v_t)$, which is the centroid position of the tessellated triangle $t$ in the parameter domain. $\omega_{t}(\eta)$ is the signed solid angle of $t$ toward $\eta$, which can be calculated by a general formulation~\citep{1983Solid}.

The coordinates of the Neumann term $\psi_{Q}^{ij}(\eta)$ can be approximately calculated as:
\begin{equation}
\begin{aligned}
\label{eq:psi_bezier_tessellation}
\psi_{Q}^{ij}(\eta) =& \sum_{t}{\iint_{(u, v) \in t}{\frac{\lambda^{ij}(u, v)}{4\pi||\textbf{b}_{t}(u,v)-\eta||}}}  \ \mathrm{d}u \mathrm{d}v \\
=& \sum_{t}{\iint_{(u, v) \in t}{\frac{\lambda^{ij}(u, v)}{||\textbf{N}_t(u, v)||} \overbrace{\frac{||\textbf{N}_t(u, v)||}{4\pi||\textbf{b}_{t}(u,v)-\eta||}\ \mathrm{d}u \mathrm{d}v}^{G(\xi, \eta) \ \mathrm{d}\xi}} }  \\
\approx & \sum_{t}{\frac{\lambda^{ij}(u_t, v_t) }{||\textbf{N}_t(u_t, v_t)||}}\int_{\xi \in t}{G(\xi, \eta) \  \mathrm{d}\xi},
\end{aligned}
\end{equation}
where $\lambda^{ij}(u_t ,v_t)$ denotes the function value of $\lambda^{ij}$ at $(u_t ,v_t)$, and $\textbf{N}_t(u_t, v_t)$ represents the unnormalized normal at $(u_t ,v_t)$. The expression $\int_{\xi \in t}{G(\xi, \eta) \mathrm{d}\xi}$ computes the integral of the fundamental solution of the 3D Laplacian equation over a planar triangle. This integral has a closed-form solution, and we directly apply the formulation given in the appendix of~\citep{2009VariationalGC}. Additionally, we incorporate the correction factor for $s(\xi)$ proposed by \citet{2022QuadGreen}, which helps to reduce the issue of the deformed object extending beyond its boundaries. The details are introduced in Appendix C.

\subsection{Linear reproduction} \label{sec:4_3}
In the previous section, we approximate the coordinates using Riemann summation. Additionally, we employ $\textbf{N}(u, v)$ of Eq.~\eqref{eq:n_bezier_point} to approximate $(\textbf{b}_u \times \textbf{b}_v)(u, v)$ in Green's third identity. As a result, the linear reproduction property (Eq.~\eqref{eq:phi_psi}) will not completely hold. Although the approximation error may not be visually noticeable in the global appearance of deformed objects, the theoretical value of this section is important, as results that fail to satisfy linear reproduction typically do not conform to the conventional theoretical definition of coordinates. We observe that \citet{2022QuadGreen} introduce a per-quad projection approach for this topic. However, this method requires precomputing a tessellated cage and projecting the coordinates of each quad onto the summation results of tessellated triangles. We believe that approximating the curved surface with piecewise linear triangles is not an optimal solution. In this work, we propose a global projection method to address this issue.

In Eqs.~\eqref{eq:phi_bezier_tessellation} and~\eqref{eq:psi_bezier_tessellation}, the approximated coordinates are derived and are now denoted as $\overline{\phi}_{Q_{k}}^{ij}(\eta)$ and $\overline{\psi}_{Q_{k}}^{ij}(\eta)$, respectively. A new index $k\in\{1,2,...,K\}$ is introduced to enumerate the Bézier patches. To ensure accurate linear reproduction, the modified coordinates $\phi_{Q_{k}}^{ij}(\eta)$ and $\psi_{Q_{k}}^{ij}(\eta)$ should satisfy the following equation precisely:
\begin{equation}
\label{eq:linear_repreduction}
\eta = \sum_{k=1}^{K}{\sum_{i=0}^{m}{\sum_{j=0}^{n}{{(\phi_{Q_{k}}^{ij}(\eta) \textbf{b}_{ij}^{Q_{k}} + \psi_{Q_{k}}^{ij}(\eta) \textbf{N}_{ij}^{Q_{k}})}}}},
\end{equation}
Additionally, we require that all  $\phi_{Q_{k}}^{ij}(\eta)$ satisfy the precise partition of unity. Specifically, the sum of all $\phi_{Q_{k}}^{ij}(\eta)$ should equal to one, because the sum of the signed solid angles of the entire closed cage is $4\pi$ with respect to $\eta$.
The above-mentioned constraints form a linear system.
\begin{equation}
\label{eq:APhi=eta}
\textbf{A} \Phi = \textbf{q},
\end{equation}
where
\begin{flalign*}
\label{eq:matrix_A}
&\mathbf{A}=
\begin{pmatrix}
\mathbf{b}_{00}^{Q_{1}} & \cdots & \mathbf{b}_{mn}^{Q_{1}} & \mathbf{b}_{00}^{Q_{2}} & \cdots & \mathbf{b}_{mn}^{Q_{K}} & \mathbf{N}_{00}^{Q_{1}} & \cdots & \mathbf{N}_{mn}^{Q_{K}} \\
1 & \cdots & 1 & 1 & \cdots & 1 & 0 & \cdots & 0
\end{pmatrix}, &\\
&\Phi = (\phi_{Q_{1}}^{00}(\eta), \cdots, \phi_{Q_{K}}^{mn}(\eta), \psi_{Q_{1}}^{00}(\eta), \cdots, \psi_{Q_{K}}^{mn}(\eta))^{\top}, \quad \mathbf{q} = (\eta, 1)^{\top}.
\end{flalign*}
Here, $\textbf{A} \in \mathbb{R}^{4 \times 2K(m+1)(n+1)}, \Phi \in \mathbb{R}^{2K(m+1)(n+1)}$, and $\textbf{q} \in \mathbb{R}^{4}$. We can prove that $\textbf{A}$ is always of full row rank, which is detailed in Appendix E. Then, we can project $\overline{\phi}_{Q_{k}}^{ij}(\eta)$ and $\overline{\psi}_{Q_{k}}^{ij}(\eta)$ onto the solution space of Eq.~\eqref{eq:APhi=eta}. Specifically, we denote 
\begin{equation*}
\label{eq:tilde_Phi}
\overline{\Phi}=(\overline{\phi}_{Q_{1}}^{00}(\eta),\cdots,\overline{\phi}_{Q_{K}}^{mn}(\eta),\overline{\psi}_{Q_{1}}^{00}(\eta),\cdots,\overline{\psi}_{Q_{K}}^{mn}(\eta))^{\top},
\end{equation*}
and formulate the optimization problem as:
\begin{equation}
\label{eq:problem}
\begin{aligned}
& \text{argmin}_{\Phi} \, ||\Phi - \overline{\Phi}||_{2}, \\
& \text{s.t.} \, \textbf{A}\Phi = \textbf{q}.
\end{aligned}
\end{equation}
Let $\textbf{u}=\Phi - \overline{\Phi}$, the problem is equivalent to finding the minimal-norm solution to $\textbf{A}\textbf{u}=\textbf{q}-\textbf{A}\overline{\Phi}$, which has the solution $\textbf{u}=\textbf{A}^{+}(\textbf{q}-\textbf{A}\overline{\Phi})$, where $\textbf{A}^{+}$ is the Moore-Penrose pseudoinverse of $\textbf{A}$. Given that $\textbf{A}$ has full row rank, its pseudoinverse can be expressed as $\textbf{A}^{+}=\textbf{A}^{\top}(\textbf{A}\textbf{A}^{\top})^{-1}$. Since  $\textbf{A}\textbf{A}^{\top}$ is a $4 \times 4$ matrix, its inverse is computationally efficient. The expression for $\Phi$ that achieves linear reproduction is:
\begin{equation}
\label{eq:new_Phi}
\Phi = \overline{\Phi} + \textbf{A}^{\top}(\textbf{A}\textbf{A}^{\top})^{-1}(\textbf{q}-\textbf{A}\overline{\Phi}).
\end{equation}

\begin{figure}[htb]
  \centering
  \includegraphics[width=0.9\linewidth]{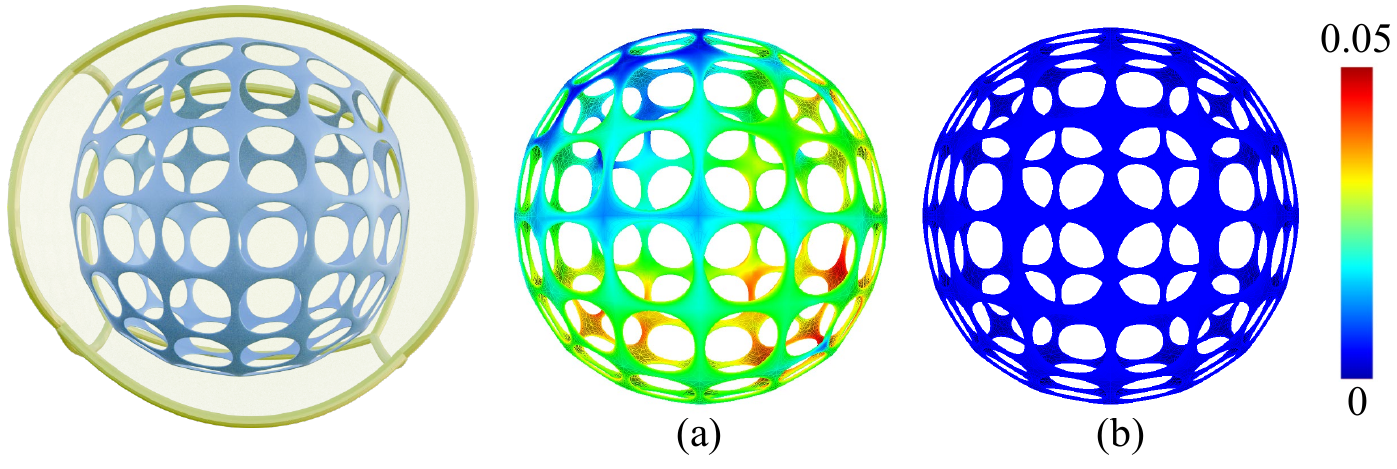}
  \caption{\label{fig:Figure0}
          Results without (a) and with (b) global projection approach. Our method perfectly achieves linear reproduction.}
\end{figure}

The most direct way to verify linear reproduction is to check if the output model matches the input model exactly when the source and target cages are identical. Fig.~\ref{fig:Figure0} (a) and (b) respectively illustrate the results without and with the global projection approach. The diameter of the sphere is about $1.94$. The color bar indicates that the Riemann summation introduces typical errors, whereas the global projection method achieves perfect linear reproduction.

\subsection{Discussion}
\begin{figure}[htb]
  \centering
  \includegraphics[width=1.0\linewidth]{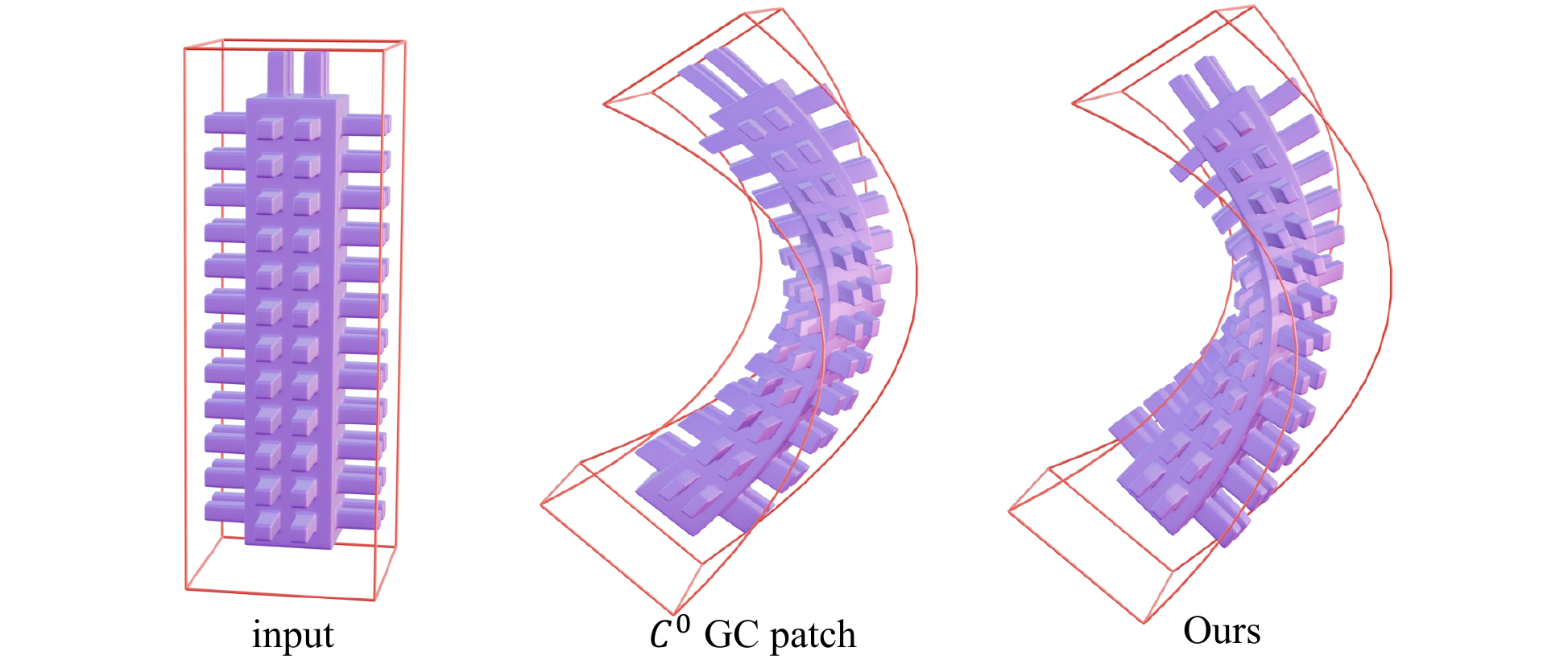}
  \caption{\label{fig:Figure8}
          Our approach demonstrates superior shape preservation properties compared to $C^{0}$ GC patch~\citep{2024GCpatych}. }
\end{figure}

In this section, we will explore the primary advantages of establishing Green coordinates for Bézier cages. A key benefit is that our method enables effective control over shape deformation. We observe that a recent work~\citep{2024GCpatych} proposes $C^{0}$ generalized Coons patches for cage-based deformation. Our method differs from this approach in the following aspects: 1) In current 3D implementation, $C^{0}$ GC patch is mainly controlled by 3D curves instead of a complete closed surface cage. 2) $C^{0}$ GC patch does not account for scenarios where the source cage is curved, as it only considers cases where the target cage is curved. This limitation restricts its ability to design a more compact source cage. 3) $C^{0}$ GC patch establishes generalized barycentric coordinates. While these coordinates are more consistent with the cage shape, the resulting deformations may exhibit shearing artifacts and appear less natural compared to the original shape. In contrast, our method integrates normal information and establishes Green coordinates with normal controls. As shown in Fig.~\ref{fig:Figure8}, our method demonstrates improved shape-preservation properties.

A degree-$(m, n)$ tensor product Bézier patch consists of $2(m+n)$ boundary control points and $(m+1)(n+1) - 2(m+n)$ interior control points. Some may be concerned that directly designing the interior control points for Bézier patches can be challenging. However, by leveraging the fact that the tensor product operation commutes with Coons operations~\citep{1992CoonsCommu}, we can directly obtain the interior control points of the bilinear blending patch from the boundary Bézier curves and construct a complete Bézier control net. This process is detailed in Appendix B. As a result, our approach does not complicate the cage design compared to $C^{0}$ GC patch. 

\section{Experiments}
\subsection{Overview}
\begin{figure}[htb]
  \centering
  \includegraphics[width=1.0\linewidth]{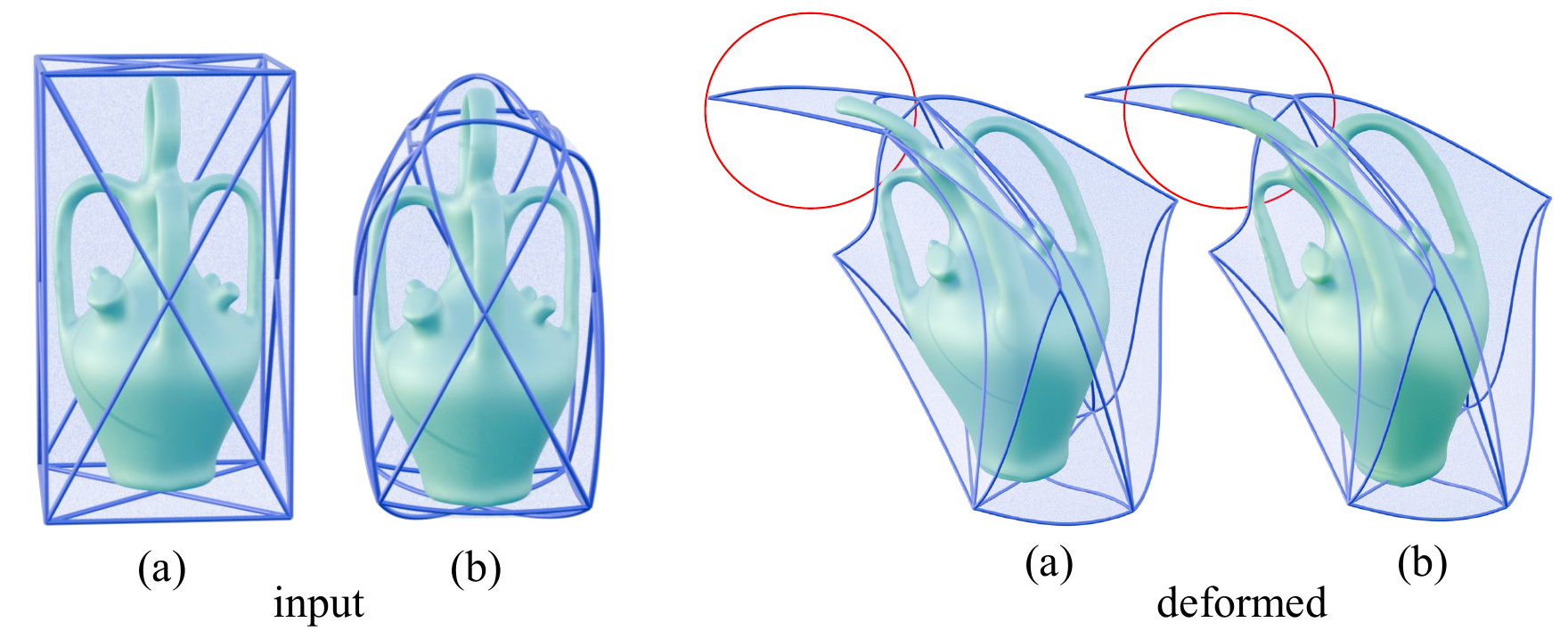}
  \caption{\label{fig:Figure3}
          Deformation results of different source cages. The tighter cage (b) yields better results. }
\end{figure}

In this section, we present the results of our method and conduct a comparative analysis with related approaches. To the best of our knowledge, our approach is the first to construct Green coordinates for 3D Bézier cages. Additionally, we did not find any publicly available Bézier cage models in previous methods. To obtain valid data for testing our algorithm, we start by modifying the quad cage provided by~\citet{2018MVCQuad}. Each quad $\textbf{q}(u, v)$ is then expanded into a degree-$(3, 3)$ Bézier patch by introducing additional control points at positions $\textbf{q}(\frac{i}{3}, \frac{j}{3})$ where $i,j\in\{0,1,2,3\}$. We subsequently design new positions for the Bézier control points, enabling the generation of compact source cages for the original shapes. The design process is completed with the help of Blender~\citep{2018Blender} and its Python script. 
Subsequently, we create the target cage by adjusting Bézier control points based on requirements, and conduct deformation using Green coordinates established on the Bézier cage. A well-designed and tightly constructed initial cage ensures that the deformed result aligns more closely with the target cage. This phenomenon is illustrated in Fig.~\ref{fig:Figure3}, where the tighter cage (b) yields superior results, as evidenced by the handle of the vase.

\subsection{Comparisons}
\begin{figure*}[htb]
  \centering
  \begin{subfigure}[b]{0.48\textwidth}
    \centering
    \includegraphics[width=\textwidth]{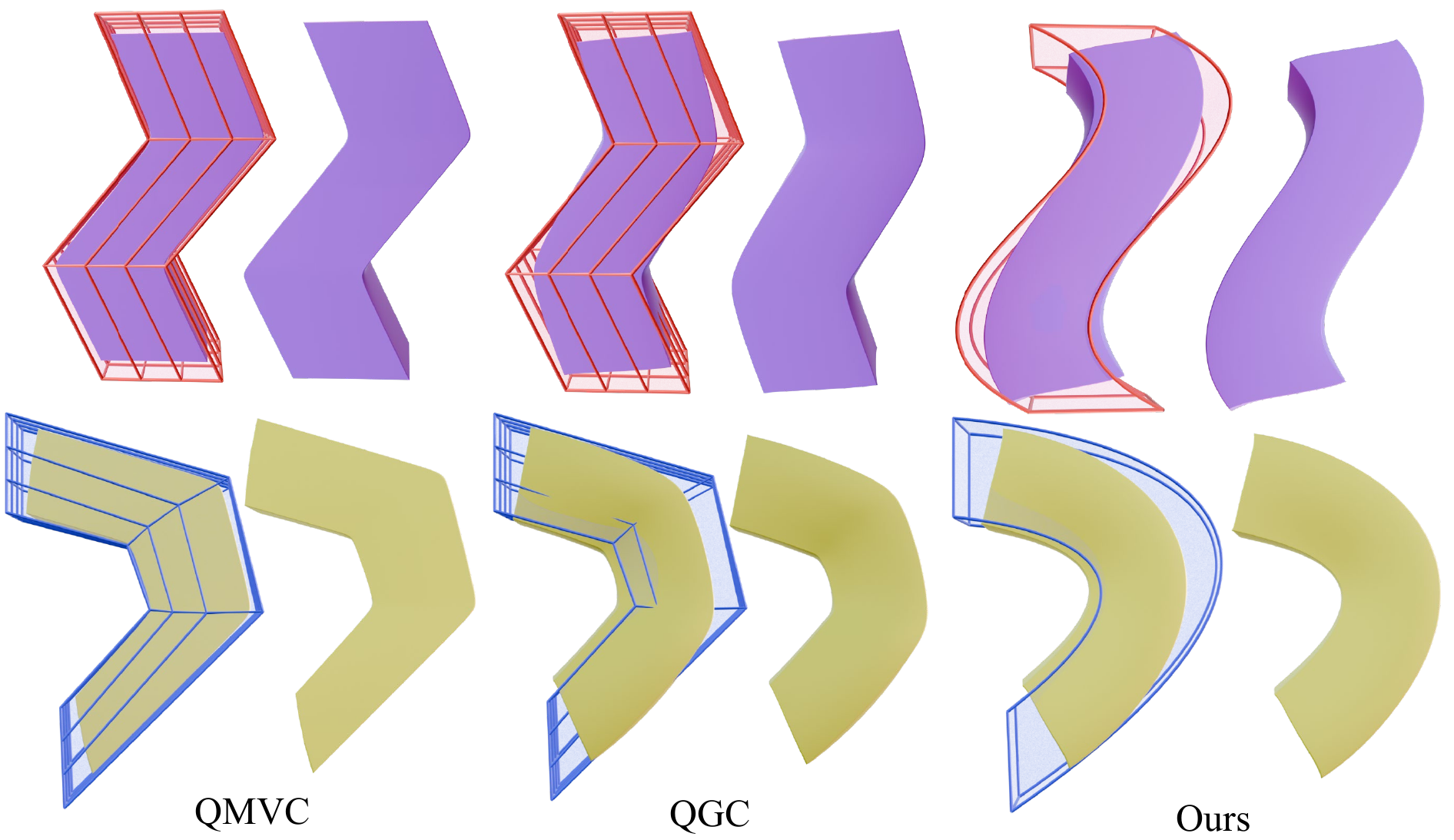}
    \caption{Comparisons of our method with QMVC~\citep{2018MVCQuad} and QGC~\citep{2022QuadGreen} on the bar model.}
  \end{subfigure}
  \hfill
  \begin{subfigure}[b]{0.48\textwidth}
    \centering
    \includegraphics[width=\textwidth]{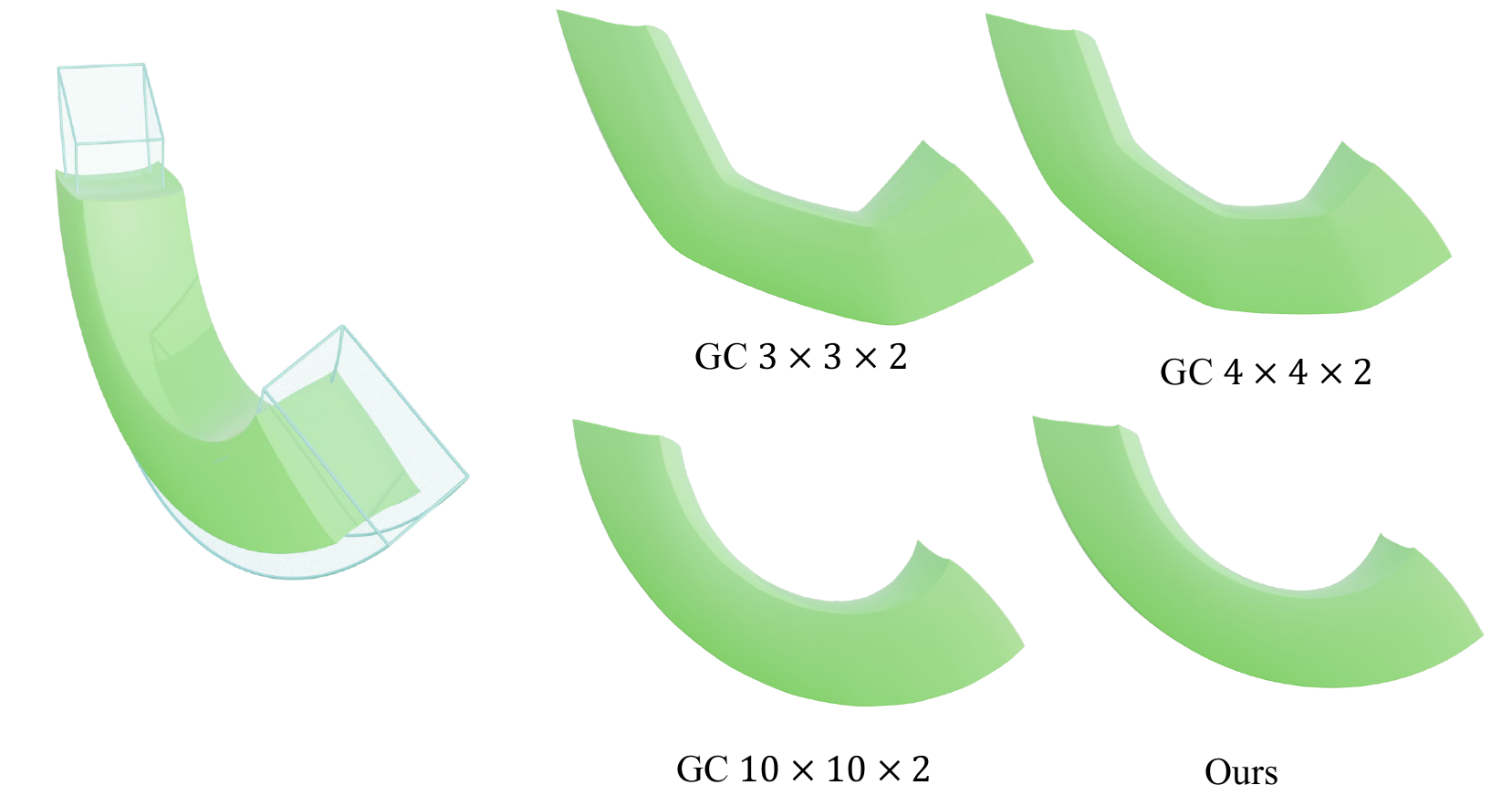}
    \caption{Comparisons of our method with GC~\citep{2008Green} on the bar model.}
  \end{subfigure}
  \caption{\label{fig:Figure9} Comparisons of our method with different methods on the bar model.}
  \label{fig:combinedFigures}
\end{figure*}

In this section, we compare our method with related approaches, including: 1) Green coordinates of triangle cages (denoted as GC hereafter)~\citep{2008Green}; 2) Mean value coordinates of quad cages (denoted as QMVC hereafter)~\citep{2018MVCQuad}; and 3) Green coordinates of quad cages (denoted as QGC hereafter)~\citep{2022QuadGreen}. We use the implementation provided by~\citet{2024Survey} for all comparative analyses. Given that quads are limited to representing straight boundaries, it is evident that using a single quad to approximate a Bézier patch would not yield good results. To make a more compelling comparison, we generate tessellated quad cages for QMVC and QGC by subdividing each degree-$(3, 3)$ Bézier patch into $9$ quads along the $u$ and $v$ directions. Specifically, for a Bézier patch $\textbf{b}(u, v)$, we introduce quad vertices at $\textbf{b}(\frac{i}{3}, \frac{j}{3})$ for $i,j\in\{0,1,2,3\}$, thereby subdividing the patch into $9$ quads. Fig.~\ref{fig:Figure9} (a) shows a typical bar model for comparison. We can observe that QMVC and QGC produces segmented structures. In Fig.~\ref{fig:Figure9} (b), we compare our method with GC, where each Bézier patch is divided into $3\times3\times2$, $4\times4\times2$, or $10\times10\times2$ triangles. It is evident that using fewer triangles still results in segmented outcomes. Although increasing the number of triangles allows for a better approximation of curved edges, it necessitates the storage of a large number of coordinates. Given that these coordinates are computed only once and do not need to be recalculated for new deformations of the same model, employing a tessellated cage would increase the time required for each deformation in Eq.~\eqref{eq:new_location}.

\begin{figure*}[htb]
  \centering
  \includegraphics[width=1.0\linewidth]{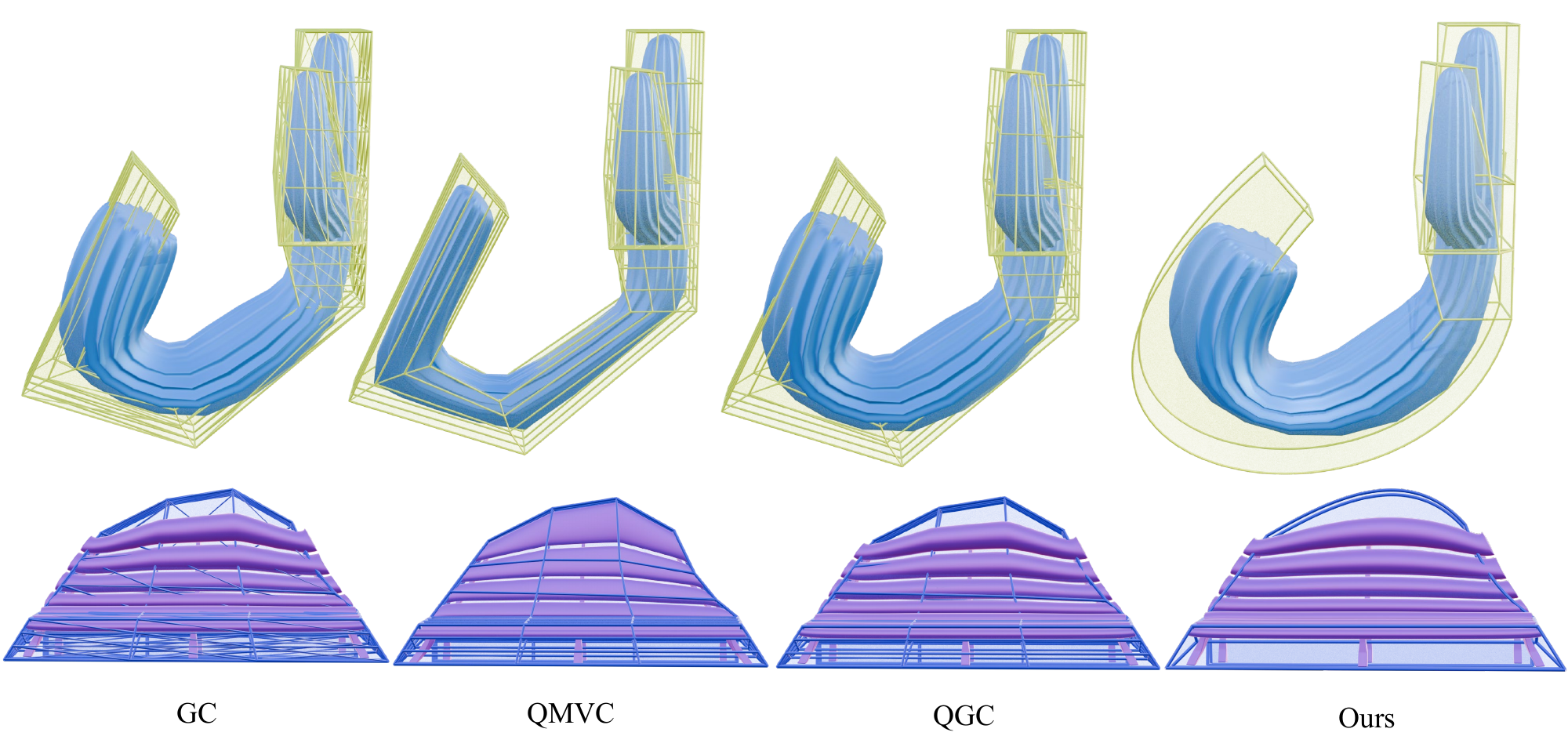}
  \caption{\label{fig:Figure4}
          Comparisons of our method with GC~\citep{2008Green}, QMVC~\citep{2018MVCQuad} and QGC~\citep{2022QuadGreen}.}
\end{figure*}
\begin{figure*}[htb]
  \centering
  \includegraphics[width=1.0\linewidth]{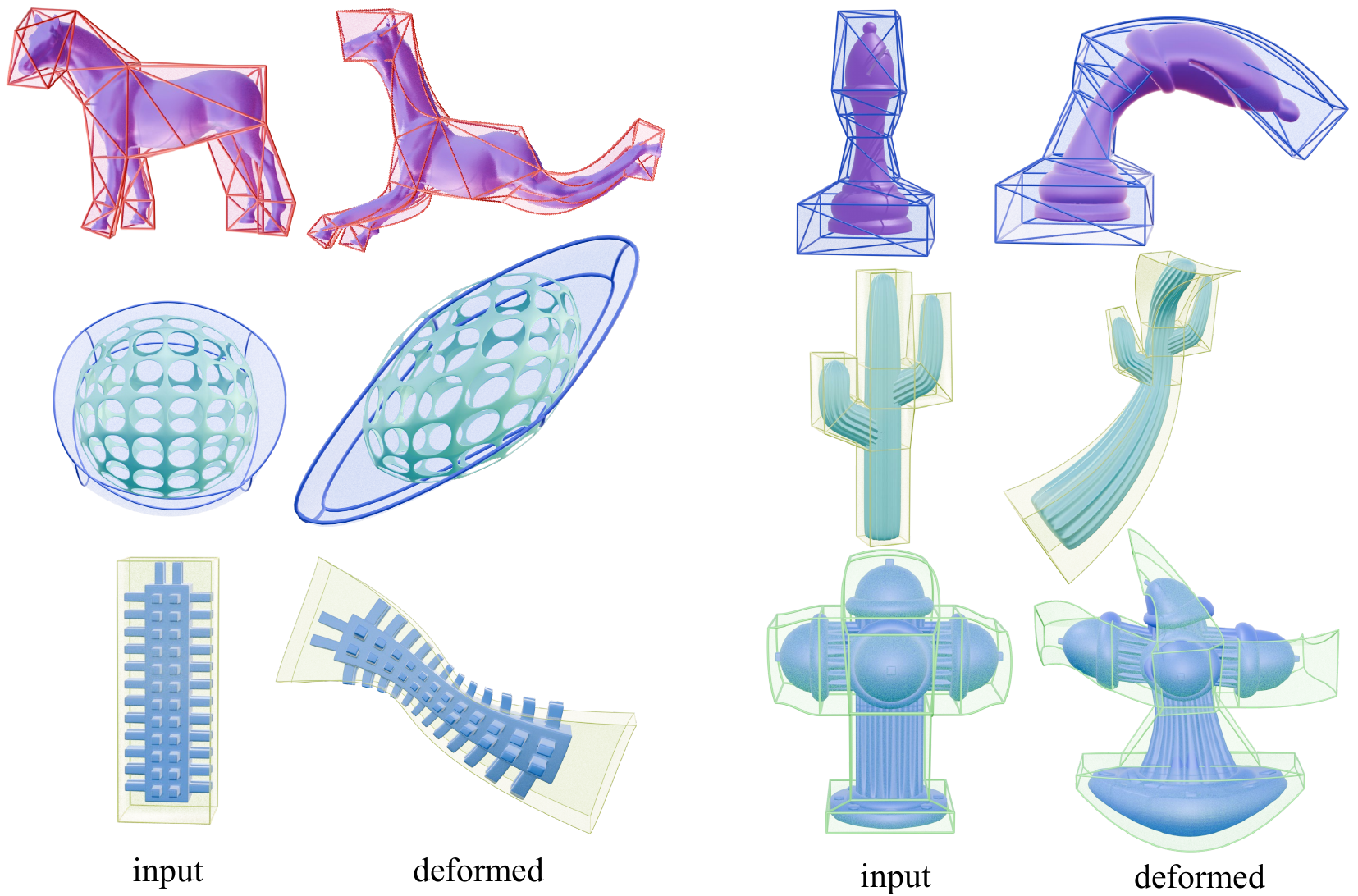}
  \caption{\label{fig:Figure11}
          Deformation results of our method in cages constructed by Bézier triangles (top row) and tensor product Bézier patches (bottom two rows).}
\end{figure*}

Fig.~\ref{fig:Figure4} presents additional comparisons with GC, QMVC, QGC, where each Bézier patch is divided into $9$ quads or $18$ triangles. We can observe that: 1) While QMVC aligns more closely with the cage, it may introduce shearing artifacts and produce less natural outputs. In contrast, Green coordinates demonstrate superior shape-preserving properties. 2) Employing Bézier cages results in better performance for bending and high-curvature deformations. When the cage is subdivided into multiple quads or triangles, it becomes evident that the deformed object exhibits segmented structures. Additionally, designing the subdivided quad cage does not appear to be easier than designing the Bézier cage. Fig.~\ref{fig:Figure11} further demonstrates the deformation results of our method applied to cages constructed by Bézier triangles (top row) and tensor product Bézier patches (bottom two rows), highlighting the broad applicability of our approach.

\begin{table}[t]
\centering
\caption{Running time of different approaches (in seconds).}
\label{table1}\resizebox{\linewidth}{!}{
\begin{tabular}{ccccccc}
\hline
Model/Method & $V$ & $B$ & GC-18 & QMVC-9 & QGC-9 & Ours\\
\hline

Cactus & 98820 & 34 & 5.00 & 102.89 & 124.62 & 26.45 \\
Bench & 65430 & 22 & 2.16 & 42.30 & 51.42 & 11.59 \\
Bar & 229378 & 6 & 2.12 & 41.57 & 49.99 & 8.51 \\
FireHydrant & 39028 & 46 & 2.67 & 54.79 & 67.04 & 14.48 \\
WireSphere & 48964 & 6 & 0.45 & 8.78 & 10.66 & 2.48 \\

\hline
\end{tabular}
}
\label{table_MAP}
\end{table}

\subsection{Running times}
Table~\ref{table1} presents the running time of different approaches. The running time of our method includes the global projection method described in Section~\ref{sec:4_3}. $V$ and $B$ represent the number of mesh vertices and the number of Bézier cage patches, respectively. All experiments are conducted on a laptop equipped with a 10-core Intel CPU at 2.40 GHz with 16GB RAM. The implementations are carried out in C++ along with Eigen libraries~\citep{2024eigen} on Windows. 

In the table, QMVC-9 and QGC-9 report the time taken by QMVC and QGC after subdividing each Bézier patch into 9 quads. Given that all quads require domain tessellation, QMVC-9 and QGC-9 require considerable computational time. GC-18 reports the time taken by GC after subdividing each Bézier patch into 18 triangles. Since GC has a closed-form solution, GC-18 is computationally efficient; however, it produces segmented shapes for high-curvature deformations. Subdividing the Bézier patch into denser triangles can better approximate the original curved cage. However, as previously mentioned, once the coordinates are pre-computed, they do not need to be recalculated for subsequent deformations. Therefore, using a tessellated cage may increase the computation time for each deformation. More importantly, this approach does not seem to be easier than designing the Bézier cage.

\section{Conclusion}
In this work, we introduce a cage-based deformation method that constructs Green Coordinates on Bézier patches, allowing for the design of tightly fitted cages and enabling flexible deformation of curved shapes. The coordinates are established on the positions and normals of the Bézier control net. Additionally, we propose a global solution-space projection method to ensure precise linear reproduction of the coordinates. Experimental results demonstrate that our method achieves smooth boundaries for bending and high-curvature deformations.

Our approach still has limitations. Firstly, the coordinates do not have a closed-form solution. Although we provide a method to ensure linear reproduction, the Riemann summation incurs a certain amount of computational time. Additionally, while Green coordinates enable shape-preserving deformations, we cannot guarantee that the deformed shape will remain entirely within the cage. In future work, we will explore closed-form solutions for high-order cages in 3D.

\bibliographystyle{ACM-Reference-Format}
\bibliography{bibliography}

\appendix
\section{Green coordinates on Bézier triangles}

This section will detail the establishment of Green coordinates on Bézier triangles. In addition to tensor product Bézier surfaces, the Bézier triangle is also a widely-used data structure for representing high-order surfaces. A degree-$n$ Bézier triangle is defined by two parameters $u$ and $v$, subject to constraints $0 \le u \le 1, 0 \le v \le 1$, and $0 \le u+v \le 1$. The parametric equation of the Bézier triangle can be expressed as:
\begin{equation}
\begin{aligned}
\label{eq:b_uv}
\textbf{b}(u,v)=\sum_{i+j+k=n \atop i,j,k\ge 0}{\frac{n!}{i!j!k!}u^{i}v^{j}{(1-u-v)}^{k}\textbf{b}_{ijk}}.
\end{aligned}
\end{equation}
We can denote $\gamma^{ijk}(u,v) = \frac{n!}{i!j!k!}u^{i}v^{j}{(1-u-v)}^{k}$ and express the above equation as:
\begin{equation}
\label{eq:abb_buvsuv2}
\textbf{b}(u, v) = \sum_{i+j+k=n \atop i,j,k\ge 0}{\gamma^{ijk}(u,v) \textbf{b}_{ijk}}.
\end{equation}

Similar to tensor product Bézier patches, we also establish Green coordinates for Bézier triangles based on the vertices and unnormalized normals of their control nets. Initially, vertex normals are constructed for each control point. Then, the normals $\textbf{N}(u, v)$ for arbitrary $(u, v)$ on the surface are computed using the coefficients of the parametric equation. The normal $\textbf{N}_{ijk}$ of the control vertex $\textbf{b}_{ijk}$ is constructed by the cross product of its one-ring neighbour. For instance, $\textbf{N}_{ijk}$ in Fig.~\ref{fig:Figure1a} is constructed by:
\begin{figure}[htb]
  \centering
  \includegraphics[width=1.0\linewidth]{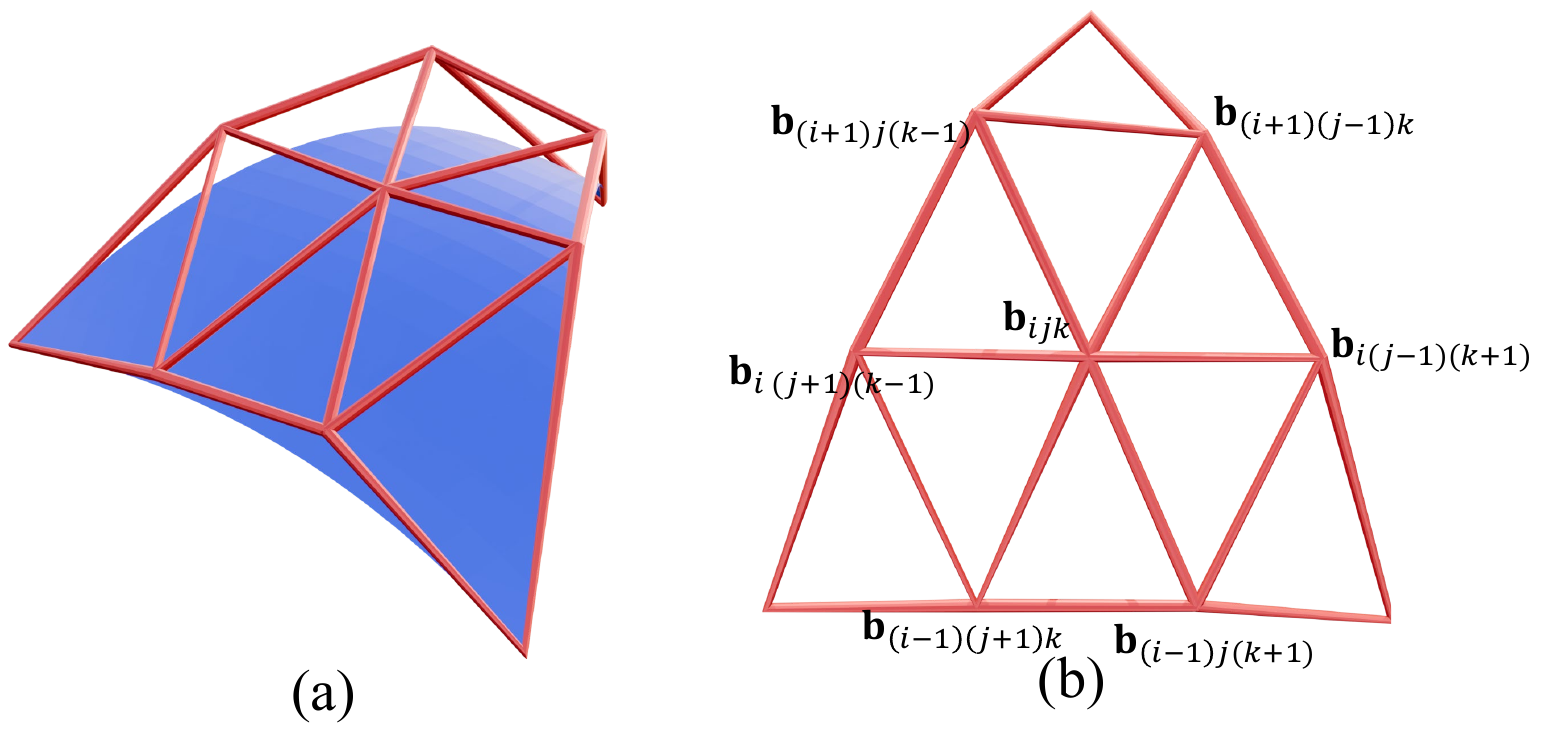}
  \caption{\label{fig:Figure1a}
          Illustration of a Bézier triangle patch (a) and its control net (b).}
\end{figure}

\begin{equation}
\label{eq:n_ij2}
\begin{aligned}
\textbf{N}_{ijk}=\frac{n^{2}}{6} &[(\textbf{b}_{(i+1)j(k-1)} - \textbf{b}_{ijk}) \times (\textbf{b}_{i(j+1)(k-1)} - \textbf{b}_{ijk}) \\
&+ (\textbf{b}_{i(j+1)(k-1)} - \textbf{b}_{ijk}) \times (\textbf{b}_{(i-1)(j+1)k} - \textbf{b}_{ijk}) \\
&+ (\textbf{b}_{(i-1)(j+1)k} - \textbf{b}_{ijk}) \times (\textbf{b}_{(i-1)j(k+1)} - \textbf{b}_{ijk}) \\
&+ (\textbf{b}_{(i-1)j(k+1)} - \textbf{b}_{ijk}) \times (\textbf{b}_{i(j-1)(k+1)} - \textbf{b}_{ijk}) \\
&+ (\textbf{b}_{i(j-1)(k+1)} - \textbf{b}_{ijk}) \times (\textbf{b}_{(i+1)(j-1)k} - \textbf{b}_{ijk}) \\
&+ (\textbf{b}_{(i+1)(j-1)k} - \textbf{b}_{ijk}) \times (\textbf{b}_{(i+1)j(k-1)} - \textbf{b}_{ijk})].
\end{aligned}
\end{equation}
Then, the surface normal $\textbf{N}(u, v)$ for parameter $(u, v)$ of the Bézier triangle is approximately calculated as:
\begin{equation}
\label{eq:n_bezier_point2}
\textbf{N}(u, v) = \sum_{i+j+k=n \atop i,j,k\ge 0}{\gamma^{ijk}(u,v) \textbf{N}_{ijk}}.
\end{equation}
$\textbf{N}(u, v)$ is linear with respect to the normals $\textbf{N}_{ijk}$ of the control vertices, which enables us to construct the Neumann term based on $\textbf{N}_{ijk}$. By applying a calculation method similar to that described in Section 4 of the main text, we can express the deformed position $\tilde{\eta}$ as:
\begin{equation}
\label{eq:phi_psi_bezier2}
\tilde{\eta} = \sum_{T}{(\tilde{f}_{D}^{T}(\eta) + \tilde{f}_{N}^{T}(\eta))} = \sum_{T}{\sum_{i+j+k=n \atop i,j,k \ge 0}{(\phi_{T}^{ijk}(\eta) \tilde{\textbf{b}}_{ijk}^{T} + \psi_{T}^{ijk}(\eta) \tilde{\textbf{N}}_{ijk}^{T})}},
\end{equation}
where
\begin{equation}
\label{eq:phi_bezier2}
\phi_{T}^{ijk}(\eta) = \int_{u=0}^{1}{\Big(\int_{v=0}^{1-u}{\frac{\gamma^{ijk}(u ,v)(\textbf{b}_{T}(u,v)-\eta) \cdot \textbf{N}_{T}(u, v)}{4\pi||\textbf{b}_T(u,v)-\eta||^{3}}}} \ \mathrm{d}v \Big) \ \mathrm{d}u,
\end{equation}
\begin{equation}
\label{eq:psi_bezier2}
\psi_{T}^{ijk}(\eta) = \int_{u=0}^{1}{\Big(\int_{v=0}^{1-u}{\frac{\gamma^{ijk}(u, v)}{4\pi||\textbf{b}_{T}(u,v)-\eta||}}} \ \mathrm{d}v \Big) \mathrm{d}u.
\end{equation}
In the above equations, $T$ represents a single Bézier triangle in the cage, $\tilde{\textbf{b}}_{ijk}^{T}$ and $\tilde{\textbf{N}}_{ijk}^{T}$ denote the positions and normals of the control vertices in the deformed cage of patch $T$, respectively.

Experimental results for Bézier triangle cages are presented in the right-hand example (the vase) of Fig. 1 and the first row of Fig. 8 (the horse and the bishop) in the main text.

\section{Generating interior vertices of bilinear blending patches}
In Section 4.4, we compare our method with $C^{0}$ GC patch~\citep{2024GCpatych} and highlight the advantage of constructing coordinates on the entire Bézier patch rather than 3D curves. Some may be concerned that directly designing interior control points for Bézier patches is quite difficult. This section mainly explains that this concern is not that necessary, as initial interior control vertices for bilinear blending patches can be easily generated when the boundary curves are provided.

Given four boundary curves, it is possible to interpolate a surface that passes through these boundaries, a concept commonly referred to as the Coons patch problem~\citep{1967Coons}. The simplest approach is bilinear blending. Considering four boundary curves that are sequentially connected, denoted as $s(0, v), s(1, v), s(u, 0), s(u, 1)$, the Coons patch $Cs(u, v)$ can be interpolated as:
\begin{equation}
\label{eq:Cs_uv}
\begin{split}
Cs(u, v) ={} & (1 - u)s(0, v) + u s(1, v) + (1 - v)s(u, 0) + v s(u, 1) \\
& - (1 - u, u) 
\begin{pmatrix}
s(0, 0) & s(0, 1) \\
s(1, 0) & s(1, 1)
\end{pmatrix}
\begin{pmatrix}
1 - v \\
v
\end{pmatrix}.
\end{split}
\end{equation}
We can verify that $Cs(u, v)$ is a parametric surface that indeed interpolates the four given boundary curves.
\begin{figure}[htb]
  \centering
  \includegraphics[width=1.0\linewidth]{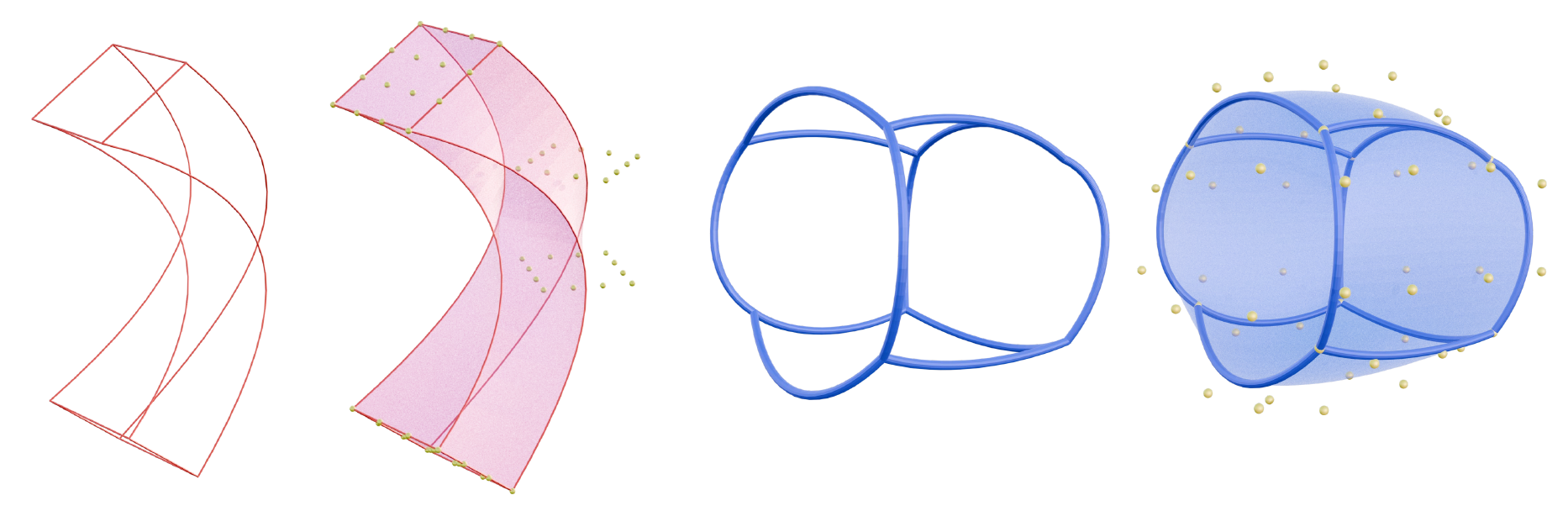}
  \caption{\label{fig:Figure_Coon}
          Given the boundary Bézier curves, we can generate the control points (the yellow-colored points) of the bilinear blending Bézier cage.}
\end{figure}
However, directly applying the Coons patch formulation to boundary curves presents a challenge in constructing Green coordinates due to the lack of interior control points, which are necessary for the construction of vertex normals. Fortunately, \citet{1992CoonsCommu} notes that the tensor product operation commutes with the Coons operation. Therefore, instead of using the formulation of the Bézier curve for $s$ as given in Eq.~\eqref{eq:Cs_uv}, we define $s(0, v)$, $s(1, v)$, $s(u, 0)$, $s(u, 1)$ as the four piecewise linear functions corresponding to the control net along the four boundary Bézier curves. Then, we position the interior control points $\textbf{b}_{ij}$ at
\begin{equation}
\label{eq:Cs_inner}
\textbf{b}_{ij} = Cs(\frac{i}{m}, \frac{j}{n}),
\end{equation}
where $i=\{1,2...,m-1\}, j=\{1,2...,n-1\}$. This process generates all the control points for the entire Bézier control net of the bilinear blending surface. Fig.~\ref{fig:Figure_Coon} illustrates several examples in which we generate the control points of the bilinear blending Bézier cage using only the boundary Bézier curves. The interior control points can be further adjusted as the user needs.

Additionally, Bézier triangles also possess a Coons patch formulation, which is illustrated in Chapter 21.6 of \citet{1993CAGDBook}.

\section{Additional details of the scale factor}
In the Green coordinates, ~\citet{2008Green} introduce a scale factor $s_j$ (Eq. (2) of the main text) to maintain rotation and scale invariance in 2D and quasi-conformality in 3D. In the continuous case, we can denote it as $s(\xi)$ or $s(u, v)$. In 2D, this factor is defined as the stretch ratio $||\tilde{t}_j||/||t_j||$ of a triangle in the cage after and before deformation. In the 3D case, the ratio that can ensure quasi-conformality is a bit more complex~\citep{2008Green, 2022QuadGreen}. For a Bézier patch $\mathbf{b}(u, v)$, the scale factor for quasi-conformality will take the following form:
\begin{equation}
\label{eq:scale_L}
s^{L}(u, v) = \sqrt{\frac{||\tilde{\textbf{b}}_u||^{2}||\textbf{b}_v||^{2} + ||\textbf{b}_u||^{2}||\tilde{\textbf{b}}_v||^{2} - 2(\tilde{\textbf{b}}_u \cdot \tilde{\textbf{b}}_v)(\textbf{b}_u \cdot \textbf{b}_v)}{2||\textbf{b}_u \times \textbf{b}_v ||^{2}}},
\end{equation}
where
\begin{gather}
\textbf{b}_{u}(u, v)=\sum_{i=0}^{m}{\sum_{j=0}^{n}{\frac{\mathrm{d} B^{m}_{i}}{\mathrm{d} u}(u)B^{n}_{j}(v) \ \textbf{b}_{ij}}}, \label{eq:bu_supp}\\ 
\textbf{b}_{v}(u, v)=\sum_{i=0}^{m}{\sum_{j=0}^{n}{B^{m}_{i}(u)\frac{\mathrm{d} B^{n}_{j}}{\mathrm{d} v}(v) \ \textbf{b}_{ij}}}. \label{eq:bv_supp}
\end{gather}
$B^{m}_{i}(u)$ and $B^{n}_{j}(v)$ are Bernstein polynomials of the tensor product Bézier patch. However, if we apply the above formulations in Eq. (15) of the main text, it is difficult to express the Neumann term based on the cage normals. Therefore, we adopt a similar method to that in Section 3.5 of~\citet{2022QuadGreen}. Specifically, during the derivation of the Neumann term, we use the following area-based ratio $s^{A}(u, v)$ instead of the aforementioned $s^{L}(u, v)$:
\begin{equation}
\label{eq:scale_A}
s^{A}(u, v) = \frac{||\tilde{\textbf{N}}(u, v)||}{||\textbf{N}(u, v)||},
\end{equation}
where $\tilde{\textbf{N}}(u, v)$ and $\textbf{N}(u, v)$ are the unnormalized normals after and before deformation at $(u, v)$. This serves as a good starting point. Subsequently, when carrying out shape deformation, we multiply the following coefficient factor $\sigma_{ij}^{Q}$ by $\psi_{Q}^{ij}(\eta)$, where
\begin{equation}
\label{eq:sigma_k}
\sigma_{ij}^{Q}=\frac{\iint_{u,v=0}^{1}{\lambda^{ij}(u, v)s^{L}_{Q}(u, v) ||\textbf{b}^{Q}_u \times \textbf{b}^{Q}_v||} \mathrm{d}u\mathrm{d}v \ }{ \iint_{u,v=0}^{1}{{\lambda^{ij}(u, v)s^{A}_{Q}(u, v) ||\textbf{b}^{Q}_u \times \textbf{b}^{Q}_v||} \mathrm{d}u\mathrm{d}v}}.
\end{equation}
Here, $Q$ represents the specific Bézier patch in the cage. Using the above coefficient factor implies that the deformed position of $\tilde{\eta}$ becomes
\begin{equation}
\label{eq:new_tilde_eta}
\tilde{\eta} = \sum_{Q}{\sum_{i=0}^{m}{\sum_{j=0}^{n}{{(\phi_{Q}^{ij}(\eta) \tilde{\textbf{b}}_{ij}^{Q} + \sigma_{ij}^{Q}\psi_{Q}^{ij}(\eta) \tilde{\textbf{N}}_{ij}^{Q})}}}}
\end{equation}
with reference to Eq. (26) of the main text. We do not explicitly write out $\sigma_{ij}^{Q}$ in the main text to avoid making the formulas overly complicated, but all $\psi_{Q}^{ij}(\eta)$ used during deformation are actually $\sigma_{ij}^{Q}\psi_{Q}^{ij}(\eta)$.

\section{Details of the cross-product Neumann term}
In Section 4.1 of the main text, we derive the Dirichlet term $\tilde{f}_{D}^{Q}(\eta)$ and the Neumann term $\tilde{f}_{N}^{Q}(\eta)$ for a single Bézier patch $Q$ as follows:
\begin{gather}
\tilde{f}_{D}^{Q}(\eta) = \iint_{u,v=0}^{1}{\frac{(\textbf{b}(u,v)-\eta) \cdot \textbf{N}(u, v)}{4\pi||\textbf{b}(u,v)-\eta||^{3}}} \tilde{\textbf{b}}(u, v)\ \mathrm{d}u \mathrm{d}v, \label{eq:f_D_supp}
\\
\tilde{f}_{N}^{Q}(\eta) = \iint_{u,v=0}^{1}{\frac{1}{4\pi||\textbf{b}(u,v)-\eta||}} (\tilde{\textbf{b}}_{u}(u, v) \times \tilde{\textbf{b}}_{v}(u, v))\ \mathrm{d}u \mathrm{d}v. \label{eq:f_N_supp}
\end{gather}
As $\tilde{\textbf{b}}(u, v)$ in $\tilde{f}_{D}^{Q}(\eta)$ represents a linear combination of the control points $\tilde{\textbf{b}}_{ij}$, the Dirichlet term of the coordinates can be established based on the positions of $\tilde{\textbf{b}}_{ij}$. Regarding the Neumann term, there are two approaches to construct the coordinates. The first is to approximate the surface normals as a linear combination of the normals of the control vertices, as shown in Eq. (25) of the main text. Another method is to utilize all cross-products of pairs of control points. Specifically, for two integer pairs $(i, j)$ and $(l, s)$ where $i,l\in\{0,1,...,m\}$ and $j,s \in \{0,1,..,n\}$, we define $(i, j) < (l ,s)$ if $i < l$ or ($i = l$ and $j < s$).  Then, we can establish the Neumann term based on all $(\tilde{\textbf{b}}_{ij} \times \tilde{\textbf{b}}_{ls})$ with $(i, j) < (l, s)$. Subsequently, the Green coordinates of the cross-product Neumann term can be written as:
\begin{equation}
\label{eq:phi_psi_bezier_cross}
\tilde{\eta} = \sum_{Q}{(\sum_{i=0}^{m}{\sum_{j=0}^{n}{{\phi_{Q}^{ij}(\eta) \tilde{\textbf{b}}_{ij}^{Q}}}} + \sum_{\substack{(i,j)=(0,0)\\(i, j)<(l,s)}}^{(m,n)}\psi_{Q}^{{ij}\times{ls}}(\eta)(\tilde{\textbf{b}}_{ij}^{Q} \times \tilde{\textbf{b}}_{ls}^{Q}))},
\end{equation}
where
\begin{equation}
\label{eq:phi_bezier_supp}
\phi_{Q}^{ij}(\eta) = \iint_{u,v=0}^{1}{\frac{\lambda^{ij}(u ,v)(\textbf{b}_{Q}(u,v)-\eta) \cdot \textbf{N}_{Q}(u, v)}{4\pi||\textbf{b}_{Q}(u,v)-\eta||^{3}}} \ \mathrm{d}u \mathrm{d}v,
\end{equation}
\begin{equation}
\label{eq:psi_bezier_supp}
\psi_{Q}^{{ij}\times{ls}}(\eta) = \iint_{u,v=0}^{1}{\frac{(\lambda_{u}^{ij}\lambda_{v}^{ls}-\lambda_{u}^{ls}\lambda_{v}^{ij})(u,v)}{4\pi||\textbf{b}_{Q}(u,v)-\eta||}} \ \mathrm{d}u \mathrm{d}v.
\end{equation}
In the above equations, $\lambda_{u}^{ij}$ is the derivative of $\lambda^{ij}$ with respect to $u$, precisely given by $\frac{\mathrm{d} B^{m}_{i}}{\mathrm{d} u}(u)B^{n}_{j}(v)$. A similar situation applies to $\lambda_{v}^{ij}, \lambda_{u}^{ls}$ and $\lambda_{v}^{ls}$.
Using this cross-product method would increase the number of Neumann term coordinates from $(m+1)(n+1)$ to $[(m+1)^{2}(n+1)^{2}-(m+1)(n+1)]/2$ compared with the method introduced in the main text. For a degree-$(3,3)$ tensor product Bézier patch, this number would increase from 16 to 120, thus raising the computation costs. The geometric meaning of the term $(\tilde{\textbf{b}}_{ij} \times \tilde{\textbf{b}}_{ls})$ for two control points is also less intuitive than that of $\tilde{\mathbf{N}}_{ij}$, which represents the normals of the control vertices. Moreover, since the influence of the Neumann term is not that sensitive or significant, the two methods produce very similar results, as shown in Fig.~\ref{fig:Figure13} below. However, the cross-product method significantly increases the running time, increasing from 26.45s to 54.93s for the Cautus model, and from 2.48s to 4.78s for the Wiresphere model.

\begin{figure}[htb]
  \centering
  \includegraphics[width=0.8\linewidth]{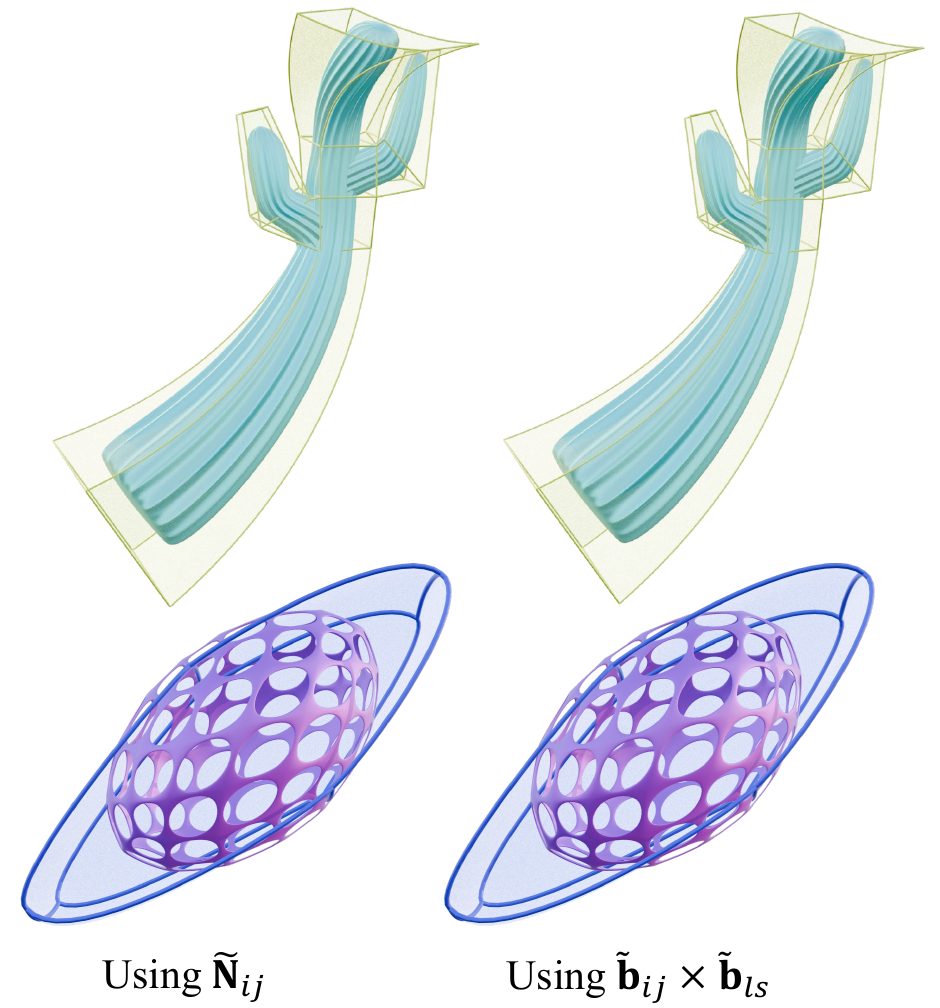}
  \caption{\label{fig:Figure13}
          Comparison between using the normal of control points $\tilde{\textbf{N}}_{ij}$ and using the cross-product of two control points $\tilde{\mathbf{b}}_{ij}\times \tilde{\mathbf{b}}_{ls}$ in the Neumann term. The two methods yield very similar results. }
\end{figure}

\section{Proof of full row rank of $\mathbf{A}$}
\begin{claim}
$\mathbf{A}$ will always have full row rank when the cage is a closed 2-manifold.
\end{claim}
\begin{proof}
Given that $\textbf{A}$ has only four rows, and considering that the row rank of $\textbf{A}$ is equal to its column rank, we only need to find $4$ linearly independent columns of $\textbf{A}$. For the closed Bézier cage, we can choose three control points $\mathbf{b}_{1}, \mathbf{b}_{2}, \mathbf{b}_{3}$ such that the three vectors $\overrightarrow{\mathbf{O}\mathbf{b}_{1}},\overrightarrow{\mathbf{O}\mathbf{b}_{2}},\overrightarrow{\mathbf{O}\mathbf{b}_{3}}$ are not coplanar, where $\mathbf{O}$ is the original point in the space. Otherwise, the entire cage will be located on a plane passing through $\mathbf{O}$, which contradicts the fact that the cage is a closed surface. Additionally, we can also find a control point $\mathbf{b}_{4}$ that does not lie in the plane formed by $\mathbf{b}_{1}, \mathbf{b}_{2}, \mathbf{b}_{3}$. Otherwise, the entire cage would lie on a plane, which also causes a contradiction. Then, if we express $\mathbf{b}_{4}$ as a linear combination of $\mathbf{b}_{1}, \mathbf{b}_{2}, \mathbf{b}_{3}$, $i.e.$,  $\mathbf{b}_{4}=x\mathbf{b}_{1}+y\mathbf{b}_{2}+z\mathbf{b}_{3}$, we will have $x+y+z\neq1$. Therefore, the four columns $(\mathbf{b}_{1},1)^{\top},(\mathbf{b}_{2},1)^{\top},(\mathbf{b}_{3},1)^{\top},(\mathbf{b}_{4},1)^{\top}$ are linearly independent, and we can prove that $\mathrm{rank}(\textbf{A}) = 4$.
\end{proof}

\end{document}